\documentclass[acmsmall,table]{acmart}
\usepackage{packages} 

\acmJournal{TOSEM} 
\acmVolume{X}
\acmNumber{X}
\acmArticle{X}
\acmYear{2025}
\acmMonth{4}
\acmDOI{10.1145/XXXXXXX.XXXXXXX} 

\AtBeginDocument{%
  \providecommand\BibTeX{{%
    \normalfont B\kern-0.5em{\scshape i\kern-0.25em b}\kern-0.8em\TeX}}}

\begin{document}

\title{Sustainability of Machine Learning-Enabled Systems: The Machine Learning Practitioner's Perspective}

\author{Vincenzo De Martino}
\email{vdemartino@unisa.it}
\orcid{0000-0003-1485-4560}
\affiliation{%
  \institution{University of Salerno}
  \city{Salerno}
  \country{Italy}
}
\affiliation{%
  \institution{Universitat Politècnica de Catalunya}
  \city{Barcelona}
  \country{Spain}
}
\email{vincenzo.de.martino@upc.edu}

\author{Stefano Lambiase}
\email{slambiase@unisa.it}
\orcid{0000-0002-9933-6203}
\affiliation{%
  \institution{University of Salerno}
  \city{Salerno}
  \country{Italy}
}

\author{Fabiano Pecorelli}
\email{fabiano.pecorelli@unipegaso.it}
\orcid{0000-0003-2446-4291}
\affiliation{%
  \institution{Pegaso University}
  \city{Naples}
  \country{Italy}
}

\author{Willem-Jan van den Heuvel}
\email{W.J.A.M.v.d.Heuvel@jads.nl}
\affiliation{%
  \institution{Tilburg University}
  \city{Tilburg}
  \country{Netherlands}
}

\author{Filomena Ferrucci}
\email{fferrucci@unisa.it}
\orcid{0000-0002-0975-8972}
\affiliation{%
  \institution{University of Salerno}
  \city{Salerno}
  \country{Italy}
}

\author{Fabio Palomba}
\email{fpalomba@unisa.it}
\orcid{0000-0001-9337-5116}
\affiliation{%
  \institution{University of Salerno}
  \city{Salerno}
  \country{Italy}
}
\renewcommand{\shortauthors}{De Martino et al.}
\begin{abstract}
Software sustainability is a key multifaceted non-functional requirement that encompasses environmental, social, and economic concerns, yet its integration into the development of Machine Learning (ML)-enabled systems remains an open challenge. While previous research has explored high-level sustainability principles and policy recommendations, limited empirical evidence exists on how sustainability is practically managed in ML workflows. Existing studies predominantly focus on environmental sustainability, e.g., carbon footprint reduction, while missing \emph{the broader spectrum of sustainability dimensions and the challenges practitioners face in real-world settings}. To address this gap, we conduct an empirical study to characterize sustainability in ML-enabled systems from a practitioner's perspective. We investigate (1) how ML engineers perceive and describe sustainability, (2) the software engineering practices they adopt to support it, and (3) the key challenges hindering its adoption. We first perform a qualitative analysis based on interviews with eight experienced ML engineers, followed by a large-scale quantitative survey with 203 ML practitioners. Our key findings reveal a significant disconnection between sustainability awareness and its systematic implementation, highlighting the need for more structured guidelines, measurement frameworks, and regulatory support.
\end{abstract}



\maketitle

\section{Introduction}
\label{sec:introduction}
Nowadays, Artificial Intelligence (AI) and Machine Learning (ML) are transformative forces that are reshaping numerous industries. These technologies effectively address complex challenges and have revolutionized fields such as healthcare, autonomous driving, and natural language processing with remarkable success \cite{ahmad2023revolutionizing,modelsaiCompaniesUsing,chib2023recent}.
Examples of AI and ML integration within real-world software systems have been showcased in both scientific and social domains~\cite{olson2011algorithm,miller2015can}, demonstrating the additional capabilities that may be provided in practice. ML, a subfield of AI, focuses on developing algorithms that learn from data~\cite{goodfellow2016deep}. In particular, AI Engineering combines machine learning and software engineering with the goal of building production-ready machine learning systems~\cite{meesters2022ai}. The increasing incorporation of ML components into traditional software systems has led to the emergence of ML-enabled systems—software systems that include at least one ML component~\cite{10.1145/3487043}. Advances in algorithms and architectures driven by researchers and tech companies have made ML-enabled systems more widespread, accurate, and performant \cite{khamparia2019systematic,shrestha2019review}.

However, significant challenges persist when deploying ML-enabled systems at scale. Training and inference processes for ML-enabled systems demand large amounts of data and substantial computational resources \cite{schwartz2020green}. Although these requirements do not prevent the adoption of ML-enabled systems, they pose serious threats to the sustainability of the systems \cite{van2021sustainable,vinuesa2020role}. Sustainable development, defined as \textit{\say{meeting present needs without compromising the ability of future generations to meet theirs}} \cite{brundtland1987our}, is a critical concern in the context of ML-enabled systems. Addressing sustainability has become a pressing issue for both practitioners and users of these systems \cite{saputri2021software,tornede2023towards}.

Software sustainability is a \textbf{multifaceted concept encompassing environmental, social, and economic dimensions} \cite{calero2013systematic,lago2015framing,venters2018software}. Environmental sustainability focuses on minimizing the carbon footprint and energy consumption of software systems, while economic sustainability ensures that software remains financially viable throughout its lifecycle. Social sustainability, in turn, refers to the extent to which software systems promote societal well-being, fairness, and inclusiveness across their development and use. This is directly tied to sustainability goals because systems that are discriminatory, opaque, or disrespectful of user rights are less likely to be trusted, adopted, or maintained in the long term. As a result, their societal and operational viability, and therefore their sustainability, might be compromised.
Rather than being a static property, sustainability in software engineering emerges from the dynamic interplay of these interconnected dimensions, as formalized by Becker et al. \cite{becker2015sustainability} in the Karlskrona Manifesto. This perspective was also remarked by McGuire et al. \cite{mcguire2023sustainability}, who defined sustainability as a \emph{stratified} and \emph{multisystemic} property, advocating for deeper analysis into how \textbf{sustainability principles evolve over time and how they are actually adopted in practice}.

Within the domain of ML-enabled systems, sustainability has become an increasingly relevant but still underexplored topic. Researchers have approached these challenges from various angles, including systematic investigations of sustainability concerns \cite{verdecchia2023systematic,tamburri2020sustainable} and the development of novel methods for integrating sustainable software engineering practices into ML workflows \cite{martinez2023towards,yarally2023uncovering}. Parallel efforts have fostered community engagement through thematic workshops, e.g., \textsc{Greens}\footnote{\url{https://greensworkshop.github.io}}, which focus on advancing discussions around sustainability-related challenges in ML.

Despite this growing academic interest, \textbf{sustainability in AI has been predominantly framed through high-level principles and policy recommendations}, rather than through concrete, actionable practices that ML practitioners can directly implement \cite{ieeeHeresCorrect,accentureMakeGenerative}. For instance, guidelines from organizations like IEEE \cite{ieeeHeresCorrect} and Accenture \cite{accentureMakeGenerative} frequently emphasize broad principles such as responsible AI, fairness, and energy efficiency but lack detailed practical methods on how to embed sustainability considerations into real-world ML development pipelines. This disconnection between principles and implementation results in sustainability being treated as an afterthought, often measured post hoc rather than integrated proactively during model design and deployment.

Moreover, the majority of sustainability-focused AI research has centered on \emph{environmental aspects}, emphasizing energy-efficient AI, carbon footprint reduction, and green software engineering practices \cite{verdecchia2023systematic,järvenpää2023synthesis}, later analyzed in open-source projects \cite{de2024developers}. While these efforts provide valuable insights, \textbf{they do not account for the broader spectrum of sustainability dimensions}, i.e., social and economic sustainability. This limited focus creates gaps in our understanding of the trade-offs practitioners face when adopting sustainability-oriented approaches in ML systems.

\steattentionboxa{\faInfoCircle \hspace{0.05cm} This paper aims to bridge the gap between high-level sustainability principles and the concrete practices employed by ML practitioners. Specifically, we conduct a mixed-method empirical investigation that systematically examines (1) how ML practitioners perceive sustainability, extending beyond theoretical perspectives to uncover industry-grounded interpretations; (2) the sustainability practices that are actively employed, identifying the approaches practitioners use to integrate sustainability considerations into ML-enabled systems; and (3) the challenges that hinder sustainability adoption in ML workflows, providing empirical evidence to inform the development of actionable frameworks, policies, and engineering solutions.}

By leveraging both qualitative insights from interviews with ML professionals and quantitative findings from a large-scale survey, this study provides a \textbf{practitioner-centric perspective on sustainability in ML}. We deliver practical recommendations, best practices, and concrete strategies that can support both researchers and practitioners in fostering sustainability across environmental, social, and economic dimensions. In doing so, this paper offers an empirical characterization of sustainability perceptions in ML, sheds light on the sustainability practices that are effectively applied, and identifies the systemic barriers that hinder their adoption.

\smallskip
\textbf{Structure of the paper.} Section \ref{sec:related} provides an overview of related work, positioning our study within the existing research on sustainability in ML-enabled systems. Section \ref{sec:methodology} describes the research design, including the research questions (\textbf{RQ}s), the mixed-method approach, and the data collection and analysis methodology. Section \ref{sec:results} presents the results, addressing how ML practitioners perceive and describe sustainability, the approaches they employ, and the challenges they face. Section \ref{sec:discussion} discusses the implications of our findings for researchers and practitioners, as well as directions for future work.
Section \ref{sec:threats} evaluates potential threats to validity and the steps to mitigate them. Finally, Section \ref{sec:conclusion} concludes the study, summarizing insights and contributions.  

\replicationpackagebox{To ensure verifiability and replicability, we have made all data and materials from this study publicly accessible \cite{appendix}. The replication package includes the complete set of interview transcripts, the coding scheme, the survey questionnaire, as well as both raw and processed survey responses and the analysis scripts. This resource validates our results and encourages future researchers to use our material for further study.}


\section{Background and Related Work}
This section provides the background and related work of our study. We begin by outlining three sustainability dimensions —social, environmental, and economic — used to analyze ML-enabled systems. These dimensions help frame sustainability as an integral component of system quality throughout the entire ML lifecycle.
Next, we survey existing literature on sustainability in ML-enabled systems, focusing on the prevailing current definition, practices, tools, and challenges. This review situates our research within the broader field of sustainable software engineering (SE) and highlights how our empirical investigation builds upon and extends prior work by empirically exploring the perspectives of ML practitioners on sustainability.


\subsection{Sustainability Dimensions in ML-enabled systems}

Sustainability, as defined in the Brundtland Report, refers to \emph{``meeting the needs of the present without compromising the ability of future generations to meet their own needs''} \cite{brundtland1987our}. While this definition originally focused on environmental and economic sustainability, it has since been applied to SE to ensure the long-term viability of software systems \cite{becker2015sustainability}. McGuire et al. \cite{mcguire2023sustainability} present a multi-layered and stratified view of sustainability, conceptualizing it as a system's capacity to influence and endure at the individual, team, organizational, and societal levels. Within this framework, sustainability is classified into three core dimensions of the software process: \emph{social}, \emph{environmental}, and \emph{economic} sustainability \cite{purvis2019three}. In the context of ML-enabled systems, De Martino et al. \cite{de2025examining} defined a theoretical framework to describe sustainable dimensions, focusing on the impact of bias mitigation algorithms. Based on this theoretical framework, these sustainability dimensions provide a comprehensive theoretical construct for understanding the impacts of ML-enabled systems. The remainder of the section overviews these three dimensions, elaborating on their relation to the overarching concept of software sustainability.

\smallskip
\textbf{Social Sustainability.} Social sustainability refers to the capacity of systems to promote equity, inclusiveness, and long-term well-being for individuals, teams, and communities involved in or affected by the development and use of technology. In the context of ML-enabled systems, this dimension includes aspects such as ethical decision-making, diversity and inclusion in design and deployment, transparency of system behavior, and respect for user privacy and autonomy.

According to McGuire et al. \cite{mcguire2023sustainability}, social sustainability can be understood at multiple levels: at the individual level, it relates to psychological well-being and cognitive effort; at the team level, it pertains to shared responsibility and ethical awareness; at the organizational level, it focuses on inclusive practices and governance structures; and at the societal level, it addresses broader issues of equity and social justice. In the context of ML development, social sustainability is reflected in the extent to which systems are designed around human-centered values, support interpretability and accountability, and actively avoid reinforcing systemic inequalities \cite{habiba2025ml}. Explainable models, for instance, serve as a key enabler by bridging communication gaps between technical and non-technical stakeholders (e.g., domain experts, regulators, or business teams), thereby promoting transparency and trust. As a consequence of these considerations, social sustainability is thus not peripheral but central to the long-term viability of software systems. ML-enabled systems that neglect fairness, inclusiveness, or accountability risk eroding user trust, impeding adoption, and facing ethical or regulatory challenges \cite{kietzmann2020deepfakes,ferrara2024fairness}. In contrast, socially sustainable systems are more likely to be accepted, maintained, and aligned with societal values over time, thereby reinforcing their overall sustainability. In our study, we adopt this perspective to capture how practitioners consider inclusiveness, responsibility, and societal impact throughout the ML lifecycle, including through education, ethical awareness, and values-aligned design decisions.

\smallskip
\textbf{Environmental Sustainability.} Environmental sustainability concerns the ability of technological systems to minimize their ecological footprint throughout their lifecycle. This  is particularly critical in the context of ML-enabled systems due to the significant computational resources required for training, inferring, and deploying ML models \cite{MLSYS2022_462211f6}. These processes often necessitate intensive hardware usage, which leads to increased energy consumption and higher carbon emissions. The environmental aspect of sustainability encompasses concerns such as energy efficiency, thermal management, resource management, and the use of hardware infrastructure that has a lower environmental impact \cite{schwartz2020green}. As ML systems become larger and more complex, their training pipelines and experimentation cycles require even more resources \cite{bender2021dangers}. Therefore, environmental sustainability encourages practitioners to adopt strategies that reduce unnecessary computations, utilize more efficient hardware or cloud-based solutions, and track environmental impact metrics such as energy consumption and CO\textsubscript{2} emissions \cite{de2025green}.

This dimension is crucial for evaluating and optimizing design decisions, from model selection to training schedules and deployment practices. In our study, we examine environmental sustainability as a lens through which ML practitioners perceive and address the ecological implications of their daily development processes. This includes practical approaches such as optimizing code and models, adopting lighter architectures, reusing pre-trained components, and monitoring resource utilization as part of an environmentally conscious decision-making process.

\smallskip
\revised{\textbf{Economic sustainability} concerns the capacity of systems to remain financially viable and resource-efficient over time. In ML-enabled systems, the substantial computational and storage demands of model development, training, and deployment often translate into high operational costs, particularly for infrastructure and personnel in large-scale or resource-limited environments \cite{bender2021dangers}. Key aspects include cost-effective development, maintainability, and efficient use of computational resources. ML teams must balance trade-offs among accuracy, efficiency, and cost \cite{gonzalez2025impact}, typically through lightweight architectures, fewer training iterations, or reuse of existing components \cite{pan2022decomposing}.}

\revised{A common way to assess the financial viability of sustainability practices is the \textit{Return on Investment (ROI)} metric, which measures profitability by comparing returns with incurred costs \cite{friedlob1996understanding}. ROI helps translate sustainability goals into clear economic terms for decision-makers. Industry tools such as the GAISSA ROI Analyzer\footnote{GAISSA Tool \url{https://gaissalabel.essi.upc.edu/}} and the Surveil AI ROI Calculator\footnote{Surveil \url{https://surveil.co/ai-roi-calculator/}} apply this concept to estimate cost savings and payback periods of optimization or green architectural strategies.}

\revised{In our study, we examine economic sustainability as a framework for understanding how practitioners assess and manage the financial implications of their work. This encompasses practices aimed at minimizing resource usage, optimizing training schedules, reducing storage requirements, and ensuring that development processes remain accessible and manageable across various organizational contexts. By analyzing these practices, we gain insights into how economic considerations influence the adoption of sustainability-oriented decisions throughout the ML pipeline.}

\subsection{Related Work}
\label{sec:related}
The rise of AI has engendered profound changes across multiple sectors, fostering advancements in sustainable products and influencing global productivity, equality, inclusion, and environmental outcomes. 
Most research explores the impact of AI in both the short and long term, emphasizing the urgent need for assessing its role in achieving the Sustainable Development Goals\footnote{The Sustainable Development Goals: \url{https://sdgs.un.org/goals}} (SDGs).\cite{vinuesa2020role,goralski2020artificial,di2020artificial}. Our research is closed aligned with the pursuit of the SDGs outlined by the United Nations. 

Venters et al. \cite{venters2018software} investigated sustainable software architectures, demonstrating their potential to enhance lifecycle management and streamline maintenance. Calero and Piattini \cite{calero2017puzzling} provided a broader perspective on software sustainability, analyzing how it has been addressed within the research community. Additionally, Oyedeji et al. \cite{oyedeji2018catalogue} introduced a Sustainability Design Catalogue, offering guidelines for integrating sustainability requirements during the design phase. Other studies have focused on environmental sustainability in SE, such as the studies by Amsel et al. \cite{amsel2011toward} and Penzenstadler \cite{penzenstadler2013towards}, which explored the direct and indirect environmental impacts of sustainable practices. Building on this knowledge, our research maps sustainability principles specifically to ML engineering practices by focusing on the unique challenges and opportunities of ML-enabled systems, and our work extends prior efforts to address the practical aspects of sustainability.

\revised{Researchers have conducted literature reviews to understand the state of the art on sustainability and Green IT, their temporal evolution, and their distribution according to research types and application domains \cite{penzenstadler2012sustainability,penzenstadler2014systematic,8379798}. These analyses revealed that sustainability in software engineering remains underexplored and has not yet influenced official standards and models. Within the Requirements Engineering (RE) community, sustainability has increasingly been framed as a non-functional requirement (NFR) that must be identified, prioritized, and balanced with other quality aspects throughout the lifecycle. The \textit{Karlskrona Manifesto} \cite{becker2015sustainability} articulated this vision by defining sustainability as a multi-dimensional concern—environmental, social, economic, individual, and technical—with both short- and long-term effects. Subsequent studies extended this conceptualization by integrating sustainability into quality models and RE activities. Venters et al. \cite{venters2018software} emphasized the need for operative definitions that connect sustainability requirements with elicitation, modeling, and trade-off reasoning, while Calero et al. \cite{calero2013systematic} proposed extensions to ISO/IEC 25010 to explicitly incorporate sustainability as a software quality characteristic. In addition, Lago \cite{8797634} introduced artifacts such as \textit{Decision Maps} to operationalize sustainability by making related concerns measurable and traceable within software design decisions. Together, these works provide a comprehensive RE-oriented foundation for operationalizing sustainability requirements. Our study builds on this foundation and extends it to the context of ML-enabled systems. While prior RE studies remained conceptual, we empirically investigate how practitioners interpret and enact sustainability requirements in ML workflows. By connecting theoretical RE perspectives with practitioner evidence, we aim to bridge the gap between the abstract framing of sustainability as an NFR and its concrete operationalization across environmental, social, and economic dimensions in ML-enabled systems.}

Groher and Weinreich \cite{8051370} and Chitchyan et al. \cite{chitchyan2016sustainability} conducted interview studies aimed at understanding how sustainability is handled in traditional software development projects. Although these studies share the same overall objective as our study, we target the peculiarities of ML-enabled systems, combining semi-structured interviews with larger-scale online questionnaires.

Some research explicitly targeted the relation between software sustainability and the development of ML-enabled systems. Aimee Van Wynsberghe \cite{van2021sustainable} proposed a formal definition of sustainable AI, highlighting the environmental costs associated with AI. At the same time, Wu et al. \cite{MLSYS2022_462211f6} examined how sustainability can be integrated throughout the entire ML model development cycle—spanning data, algorithms, and hardware—to mitigate AI's overall carbon footprint, providing principles for optimizing resource utilization, improving energy efficiency, and reducing environmental impact. With respect to these papers, ours presents two main differences. First, we aim to understand how ML practitioners perceive in ML systems, in an effort to elicit the elements that they perceive as most important for the development of sustainability ML-enabled systems. Second, our work aims to map the concept of sustainability into practices that can be used to build sustainable products, encompassing the social, environmental, and economic dimensions of sustainability. Last but not least, from a conceptual perspective, our mixed-method research approach aims to merge the interdisciplinary perspective among AI, SE, and sustainability, expanding previous research efforts with socio-technical approaches. 

More recently, McGuire et al. \cite{10172842} have improved sustainability theories in SE, calling for further research on how to instill sustainability principles in software development processes and products. Our work goes toward this direction by examining sustainability in ML-enabled systems and incorporating socio-technical perspectives.

Finally, Järvenpää et al. \cite{järvenpää2023synthesis} reviewed green tactics proposed in the literature \cite{verdecchia2023systematic} and evaluated their application in a focus group with three ML practitioners, and later De Martino et al. \cite{de2024developers} investigated their adoption in open-source projects. Our work must be seen as complementary for three reasons. In the first place, we did not focus on the practices defined in the literature, but we elicited sustainability approaches by inquiring practitioners. The application of this research method has two implications. On the one hand, we could assess whether the practices defined in the literature are also employed in practice, hence assessing the impact of the solutions proposed by the SE research community so far. On the other hand, we could also extend the array of practices to build sustainable ML-enabled systems, thus informing SE researchers of the next steps to pursue. In the second place, our work has a broader landscape. While Järvenpää et al. \cite{järvenpää2023synthesis} mainly focused on the practices that impact \emph{environmental} sustainability, we also considered additional dimensions such as \emph{social} and \emph{economic}. In this sense, our work could therefore provide a more comprehensive overview of how practitioners deal with sustainability in ML-enabled systems. Last but not least, our work does not only aim at eliciting the approaches used by ML practitioners when building sustainable ML-enabled systems, but also at (1) investigating the perception of sustainability in practice and (2) cataloging the challenges faced by practitioners. 

\novelty{Unlike previous research that focuses predominantly on high-level principles or environmental aspects, our work adopts mixed-method research to provide a practitioner-centered perspective. By analyzing sustainability across environmental, social, and economic dimensions, we uncover real-world practices and challenges faced by ML engineers. Furthermore, this study contributes novel insights into sustainability-driven engineering practices and provides recommendations to bridge the gap between theoretical principles and practical adoption in ML workflows.}

\section{Research Process Overview}
\label{sec:methodology}

The study aimed to explore how machine learning (ML) practitioners involved in developing ML-enabled systems approach sustainability aspects. Specifically, the study focused on (1) understanding how ML practitioners perceive and describe sustainability, (2) revealing the sustainability practices they actively use, and (3) identifying the challenges that hinder the integration of sustainability in ML workflows. To achieve the above-mentioned objective, we delineated three main research questions (one for each objective). 

First, we aimed to understand how ML practitioners describe and interpret sustainability in ML-enabled systems. Such a preliminary investigation allowed us to challenge the practitioners' understanding of the concept, potentially uncovering alternative perspectives, misconceptions, or gaps in their reasoning. By analyzing their descriptions, we identified areas of consensus and key points of divergence in how sustainability is perceived within the ML community. Additionally, we could compare these perceptions with the definitions of sustainability proposed by researchers to highlight potential alignments and discrepancies.

\sterqbox{RQ\textsubscript{1}—Sustainability Perception}{How do machine learning practitioners perceive sustainability in ML-enabled systems?}

The second research question sought to understand the practices—including tools and measures—applied by ML practitioners to address ML sustainability. As reported earlier in the paper, previous major research efforts have primarily focused on identifying green and environmental practices, creating catalogs of green AI approaches \cite{verdecchia2023systematic,järvenpää2023synthesis}. In the context of our work, we aim to extend the current body of knowledge by considering a more comprehensive definition of sustainability, which includes aspects such as social and economic sustainability \cite{purvis2019three}. This knowledge could inspire researchers and practitioners, providing insights for configuring and validating automated approaches to improve system sustainability by reducing design errors. They can also help establish best practices, standards, and guidelines that foster the development of sustainable ML. By sharing these strategies and their impacts, we hope to encourage widespread adoption and discussions within the community.

\sterqbox{RQ\textsubscript{2}—Sustainability Approaches}{What approaches, metrics, and tools do machine learning practitioners employ in building sustainable ML-enabled systems?}




Finally, our research aimed to highlight ML practitioner's challenges while working on sustainability. Previous research has primarily examined traditional software \cite{8051370,chitchyan2016sustainability}, neglecting the specific challenges associated with ML-enabled systems. This perspective can serve both researchers and practitioners in designing novel solutions and enhancing the current state of practice.

\sterqbox{RQ\textsubscript{3}—Sustainability Challenges}{What are the challenges faced by machine learning practitioners when building sustainable ML-enabled systems?}

To address our research questions, we applied a \textit{mixed-method} research approach~\cite{johnson2007toward} with two main phases. The structure of our paper was based on a similar study conducted by Al-Subaihin et al. \cite{al2019app}, which employed a mixed-method approach using interviews and surveys. The first phase is an explorative \emph{qualitative} study; we conducted semi-structured interviews—following established guidelines~\cite{hove2005_semi-structured_interviews}—with eight practitioners and analyzed the collected data using \textit{Socio-Technical Grounded Theory (STGT) for data analysis}~\cite{9520216} incorporating techniques such as open coding, constant comparison, and memoing. Our focus during this analysis was on uncovering and explaining socio-technical phenomena related to ML practitioners' work practices regarding sustainability \cite{uchitel2024scoping}. This strategy was adopted based on the consideration that the integration of sustainability aspects into software development is a complex phenomenon intrinsically linked to human factors. For this reason, a purely quantitative approach would not have been sufficient to capture all relevant aspects. Consequently, a qualitative approach was chosen, as it allows greater flexibility and adaptability throughout the research process, enabling adjustments based on emerging findings~\cite{razali2023combining, 9520216, strauss1997_grounded_theory, strauss1990_basics_of_qualitative_research, glaser2017_original_grounded_theory}.

However, the insights from the initial qualitative phase were limited in scope and not necessarily representative of a broader population. To address this limitation, we conducted a second \emph{quantitative} phase using an online survey. This approach allowed us to gather responses from a more diverse sample—spanning various demographics, professional roles, and geographic locations. To conduct this study, we followed the guidelines established by Dillman et al. \cite{dillman2014internet} and Kitchenham \cite{kitchenham2008_PersonalOpinionSurveys} for survey studies. The large-scale survey provided quantitative insights that complemented our qualitative findings, thereby enhancing both the robustness and validity of our research. In this way, the survey served as an additional step, ensuring that our conclusions were grounded in both the in-depth qualitative insights and the broader quantitative validation.

\begin{figure}
    \centering
    \includegraphics[width=1\linewidth]{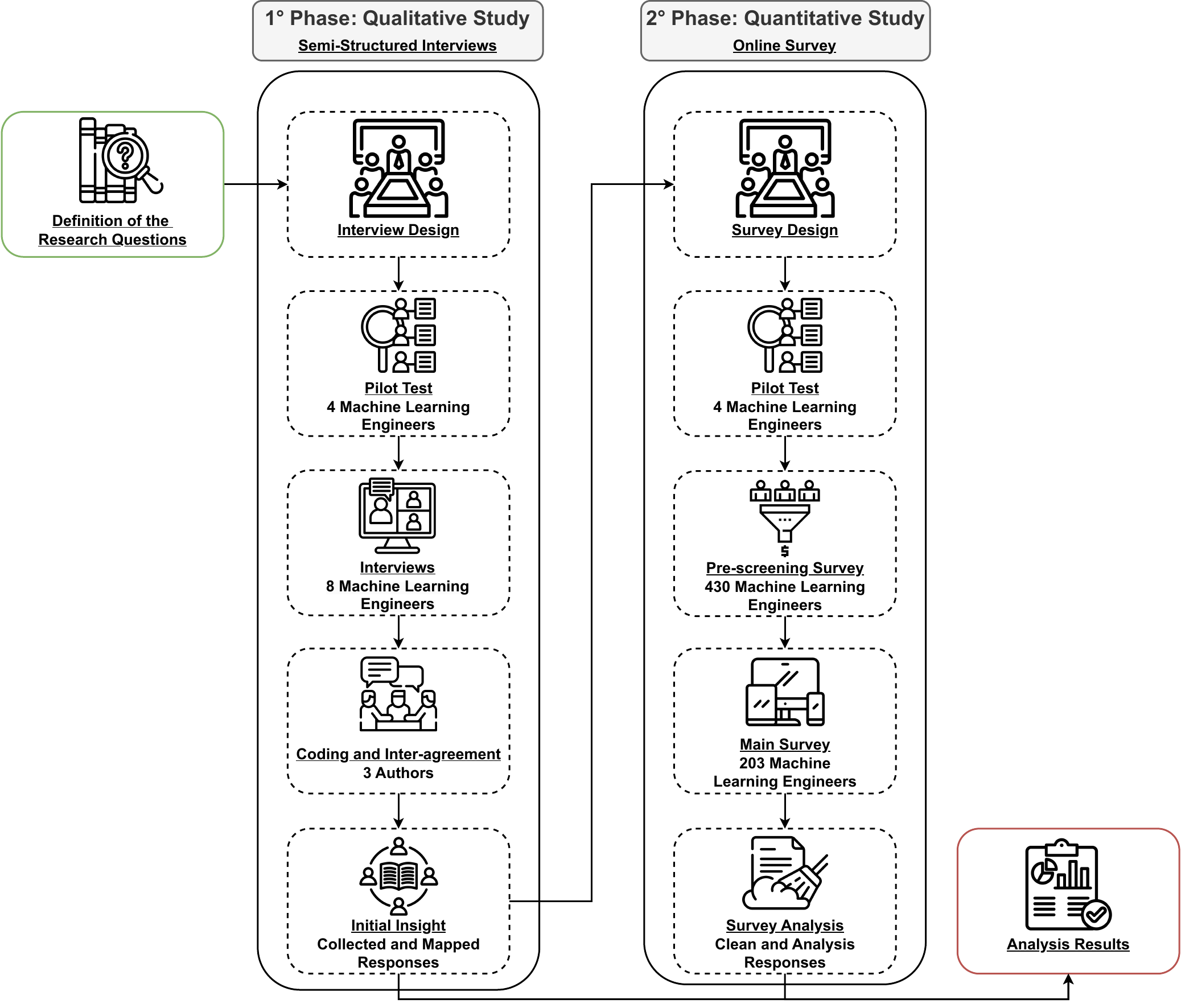}
    \caption{Overview of the research process.}
    \label{fig:overview}
\end{figure}

For the result reporting, we employed the guidelines by Wohlin et al.~\cite{wohlin2012experimentation} and Hoda~\cite{9520216}, following the ACM/SIGSOFT Empirical Standards.\footnote{Available at: \url{https://github.com/acmsigsoft/EmpiricalStandards}. Given the nature of our study and the currently available standards, we followed the \textsl{``General Standard''}, \textsl{``Questionnaire Surveys''}, \textsl{``Qualitative Surveys''}, and \textsl{``Mixed Methods''} guidelines.} In the reminder of this section, we describe our overall study design, as depicted in Figure \ref{fig:overview}.

\begin{table}[ht]
    \caption{Structure and examples of questions of semi-structured interviews.}
    \label{table:question}
    \scriptsize
    \begin{tabular}{|p{0.45\linewidth}|p{0.5\linewidth}|} \hline
    \rowcolor{black}
    \textcolor{white}{\textbf{Question}} & 
    \textcolor{white}{\textbf{Rationale}} \\ \hline
    \rowcolor{gray!15}
    \multicolumn{2}{|c|}{\textbf{Part 1: Introduction and Welcome}} \\ 
    \hline
    What is your nationality? & Understanding participants' nationality helps contextualize their perspectives on sustainability in ML-enabled systems. \\
    \hline
    What is your highest academic degree? & Understanding participants' academic background helps contextualize their perspectives on sustainability in ML-enabled systems. \\
    \hline
    How would you describe your company’s business? & Identifying the industry sector allows for the analysis of sustainability approaches in different organizational contexts. \\
    \hline
    How many employees does your company have? & Company size can influence the adoption of sustainability practices and the availability of resources to implement them. \\
    \hline
    What is your corporate role? & The role of the participant can affect their involvement in sustainability-related decisions within ML projects. \\
    \hline
    How many years have you worked in this role? & Experience level provides insight into whether sustainability considerations evolve with industry tenure. \\
    \hline
    What is your work field? & The specific work domain helps assess how sustainability challenges vary across different ML applications. \\
    \hline
    \rowcolor{gray!15}
    \multicolumn{2}{|c|}{\textbf{Part 2: Sustainability Perception (RQ$_1$)}} \\
    \hline
    How do you perceive the concept of sustainability? Particularly from a "Social", "Environmental" and "Economic" point of view. & This question helps explore how ML practitioners conceptualize sustainability across multiple dimensions. \\
    \hline
    Among the sustainability domains, which one (out of the three defined above) do you attach the most value to during the development of an ML project? & Identifying the most valued sustainability dimension provides insights into prioritization biases among ML practitioners. \\
    \hline
    Can you sort the sustainability dimensions based on your importance? & Understanding the relative importance of sustainability aspects helps contextualize practitioners’ decision-making processes. This ranking question was included to mitigate bias by forcing a prioritization decision among dimensions. \\
    \hline
    \rowcolor{gray!15}
    \multicolumn{2}{|c|}{\textbf{Part 3: Sustainability Approaches (RQ$_2$)}} \\
    \hline
    What tools and metrics do you adopt to improve sustainability aspects? & This question identifies the specific tools—and associated metrics—used to incorporate sustainability into ML development. \\
    \hline
    What methodologies do you adopt to improve sustainability aspects? & Investigating methodologies provides insight into structured processes for embedding sustainability into ML workflows. \\
    \hline
    What frameworks do you adopt to improve sustainability aspects? & Understanding the frameworks practitioners rely on can highlight existing best practices and gaps in sustainability integration. \\
    \hline
    How does the development of ML systems change by adopting the described practices? & Examining changes introduced by sustainability practices allows for evaluating their real-world impact on ML development. \\
    \hline
    Once you have adopted these sustainable practices, do you notice a decrease in the accuracy or overall quality of the models? & This question investigates potential trade-offs between sustainability and model performance, a key concern in ML adoption. \\
    \hline
    What are the main trade-offs of these approaches? & Understanding trade-offs allows for the identification of barriers that might prevent practitioners from adopting sustainability measures. \\
    \hline
    \rowcolor{gray!15}
    \multicolumn{2}{|c|}{\textbf{Part 4: Sustainability Challenges (RQ$_3$)}} \\
    \hline
    What are the main challenges you face when developing sustainable ML systems? & Identifying primary obstacles helps outline key barriers to sustainability adoption. \\
    \hline
    What are the main challenges you face in integrating these approaches and how successful are they in improving sustainability in the ML systems? & Evaluating challenges in integration helps determine whether sustainability initiatives achieve their intended impact. \\
    \hline
    Could you tell if you have any additional problems, needs or other issues when dealing with the development of sustainable ML systems? & Exploring additional concerns can uncover previously unaddressed barriers in sustainability adoption. \\
    \hline
    Could you share some examples of challenges and how they were addressed? & Case studies and practical experiences provide concrete insights into overcoming sustainability-related obstacles. \\
    \hline
    \end{tabular}
\end{table}

\subsection{First Phase: Qualitative Study}

As highlighted at the beginning of this section, the first step in the research process focused on gaining an in-depth understanding of the phenomenon through a qualitative approach. As already mentioned, we employed \textit{STGT for data analysis} method proposed by Hoda~\cite{9520216}. This method emphasizes data collection techniques that yield natural language text, which can be effectively obtained through semi-structured interviews. The following section provides a detailed description of these methodological aspects.

\subsubsection{Design of the Semi-structured Interview}

The questions for the semi-structured interviews were developed according to the guidelines established by Hove and Anda~\cite{hove2005_semi-structured_interviews}.
We used semi-structured interviews because they blended predefined questions (designed to explore the primary research topic) with open-ended inquiries (intended to uncover unforeseen insights). This qualitative approach is commonly employed to collect in-depth insights, perspectives, and personal anecdotes from participants \cite{creswell2016qualitative}, all while ensuring consistency and comparability across interviews.
The first author initially created the interview design and then refined it collaboratively with the other authors during recurrent meetings. It is organized into four parts, as illustrated in Table \ref{table:question}.

The interviews followed a funnel model, beginning with broad questions and gradually moving toward more specific topics. This approach facilitated a natural conversational flow, allowing participants to speak freely before the interviewer guided the discussion to more focused areas \cite{al2019app}. At the outset, participants were informed about the study’s primary focus and objectives, as well as the following conditions: (1) their participation was voluntary; (2) they were free to withdraw from the study at any time; and (3) their responses would be aggregated and used for a scientific publication. Once the participants agreed to these terms, we proceeded with the interview.

Before conducting the interviews, the first author performed a pilot study with four ML practitioners \cite{meesters2022ai}. This phase was crucial for identifying potential issues in the interview process, such as unclear questions or challenges in maintaining a natural flow of conversation. The pilot participants, each with at least two years of experience in ML, confirmed the quality of the structure and content of the interview.
Additionally, we assessed the duration of the pilot interviews to ensure they were appropriately brief.

\subsubsection{Structure of the Semi-strucured Interview}

As is well known, semi-structured interviews involve a general set of questions designed to initiate discussion with the participant, which can (and should) be tailored based on the specific context. In particular, the interview approach followed a structured strategy consisting of four key phases. Table \ref{table:question} presents a selection of these questions, generalized across all conducted interviews.

\begin{enumerate}
    \item \textbf{Introduction and Rapport Building:} The participant was welcomed and made comfortable to establish a rapport. The interviewer(s) and the interviewee initially introduced themselves. Afterwards, technical details were provided, and basic demographic information was collected.

    \smallskip
    \item \textbf{Understanding Sustainability:} Immediately afterward, in line with the first research question, the participant was introduced to a definition of sustainability encompassing all relevant dimensions. They were then asked to share and discuss their personal understanding of sustainability through an open-ended question—in this way, we collected unprompted and spontaneous views on the topic. This initial step was purposefully designed to mitigate potential biases such as social desirability or acquiescence \cite{grimm2010social}, as it allowed participants to express their thoughts freely without being influenced by predefined categories or response options. Only after this exploratory question, did we introduce two structured formats: (i) a Likert-scale assessment where participants rated the importance of the three sustainability dimensions (environmental, social, and economic), and (ii) a ranking task where they were asked to order these dimensions based on perceived importance. The combination of these formats enabled both a broad understanding and a prioritization of sustainability concerns.

    \smallskip
    \item \textbf{Sustainability Approaches in Practice:} The third phase focused on the second research question, where participants were asked about the approaches adopted in their workplace to support sustainability. This included inquiries about their awareness and use of tools, as well as whether they considered sustainability-related metrics for evaluation. Additionally, participants were asked whether these approaches had a tangible impact.

    \smallskip
    \item \textbf{Challenges in Sustainable Software Development:} In the fourth phase, the discussion shifted to the final research question, addressing the challenges encountered when attempting to support sustainability or, more broadly, when dealing with this topic in software development. Participants were asked to reflect on the key challenges they typically encounter when addressing sustainability in practice and to share real-world examples of the difficulties they face in integrating sustainability principles into ML-enabled systems. This approach enabled us to gain practical insights into the constraints, trade-offs, and decision-making processes that shape sustainable ML development.

    \smallskip
    \item \textbf{Conclusion and Final Remarks:} Finally, the interview concluded with a farewell, giving participants the opportunity to ask any remaining questions and receive a recap of key information relevant to them. They were sincerely thanked for their time and effort in contributing to the discussion. 
\end{enumerate}

\subsubsection{Interview Sample and Procedure}

As for the participant selection process, we opted for a convenience sampling method~\cite{hair2007_convenience_sampling_def,baltes2022_convenience_sampling_SE}. This approach involves non-probabilistic sampling, where individuals are chosen based on proximity, availability, or willingness to participate in the research. However, it is worth noting that this method does present limitations regarding the generalizability of findings~\cite{hair2007_convenience_sampling_def}. For such a reason (other than conducting a survey study described in Section \ref{sec:survey}), we identified a set of criteria that our participants had to meet.  
\begin{itemize}
    \item We targeted Machine Learning Practitioners, \ie ML Engineers, Data Scientists, Artificial Intelligence Engineers, and Software Engineers who primarily focus on producing ML software~\cite{meesters2022ai,bosch2021engineering}. 
    \item These practitioners were required to be employed at organizations actively involved in ML-enabled system development and demonstrate an interest in sustainability.
\end{itemize}

\begin{table}[ht]
    \caption{Information of the interview study participants.}
    \label{table:participants}
    \scriptsize
    \rowcolors{1}{gray!15}{white}
    \begin{tabular}{|c|c|c|c|c|c|c|c|} \hline
    \rowcolor{black}
    \textcolor{white}{\textbf{ID}} & 
    \textcolor{white}{\textbf{Nationality}} & 
    \textcolor{white}{\textbf{Years of Exp.}} & 
    \textcolor{white}{\textbf{Role}} & 
    \textcolor{white}{\textbf{Company Size}} & 
    \textcolor{white}{\textbf{Degree}} & 
    \textcolor{white}{\textbf{Work Domains}} \\ \hline
    P1 & Italy & 4 & ML Engineer & 0-10 & Master's Degree & Medical and Scientific \\ \hline
    P2 & Brazil & 8 & ML Engineer & 50-200 & Master's Degree & Economic and Political \\ \hline
    P3 & Italy & 6 & ML Engineer & 0-10 & Master's Degree & Medical and Scientific, Technical \\ \hline
    P4 & Italy & 1 & Data Scientist & Over 500 & Master's Degree & Socio-Cultural \\ \hline
    P5 & Italy & 3 & Data Engineer & 250-500 & Bachelor's Degree & Economic and Political \\ \hline
    P6 & Italy & 1 & AI Engineer & Over 500 & Master's Degree & Economic and Political \\ \hline
    P7 & Switzerland & 5 & Software Engineer & Over 500 & Master's Degree & Technical \\ \hline
    P8 & Germany & 5 & Software Engineer & 250-500 & Master's Degree & Economic and Political \\ \hline
    \multicolumn{7}{p{1\linewidth}}{\tiny Notes on the values in the last column: Medical and Scientific: applies ML in psychology, biology, or medical research; Economic and Political: focuses on finance, economy, or politic fields;  Technical: focuses on information systems, transportation or urbanistic; Socio-Cultural: uses ML for social trends, education, or sports;}
    \end{tabular}
\end{table}

Starting from a candidate set of 35 ML practitioners from our network, we conducted interviews in an iterative approach of data collection and analysis, until reaching theoretical saturation~\cite{saunders2018saturation}, \ie when no new subtopics are identified in different interviews. In the end, 8 interviews were jointly conducted by all the authors of the paper (Table \ref{table:participants} reports the participant characteristics). As shown, most participants have a rather long experience with the development of ML-enabled systems and work for medium-large companies having different business goals. As such, they could provide a diverse set of insights into the sustainability practices employed in different contexts. 

The interviews were conducted through \textsl{Google Meet}\footnote{\url{https://meet.google.com/}} and \textsl{Microsoft Teams},\footnote{\url{https://www.microsoft.com/microsoft-teams/group-chat-software?rtc=1}} depending on the instrument preferred by the practitioners. On average, the interviews took for 36 minutes, ranging from a minimum of 19 to a maximum of 55 minutes.

\subsubsection{Data Analysis}

We recorded the interviews and extracted a transcription of them. On such data, we applied \textit{STGT for data analysis}~\cite{9520216}. This approach includes techniques such as open coding, constant comparison, and memoing, which facilitated structured but flexible data analysis. 

The transcribed data and collected information were stored in a CSV file for analysis. The coding process began with a first phase of \textit{open coding} and \textit{in-vivo coding}, aimed at identifying and labeling significant segments of text based on participants' information and recurring themes. In this phase, the first three authors independently analyzed separate transcripts to ensure initial diversity in perspectives. In the second phase, \textit{conceptual coding} was conducted to organize the identified codes into broader concepts. The three authors reviewed each other’s coding work and met to discuss and reconcile any discrepancies. Disagreements were systematically addressed through discussion, during which authors presented their interpretations and sought consensus. To ensure rigor, we employed Cohen's Kappa~\cite{cohen1960coefficient} as an inter-rater reliability measure. Each transcript's coding was assessed on a 0–5 scale to validate agreement among the authors, with 0 indicating \textit{Strongly disagree} and 5 \textit{Strongly Agree}. A Cohen’s Kappa value of 0.75 or higher was considered acceptable, indicating substantial agreement. Once consensus was reached, the concepts were further grouped into categories, representing higher-level themes. The categorization process was iterative and relied on group discussions to refine the findings, minimizing subjective bias. 

To maintain transparency and reproducibility, all details of the coding procedure and resulting categorizations are available in the online appendix~\cite{appendix}.


\subsection{Second Phase: Quantitative Survey}
\label{sec:survey}

As a second step of our methodology, we operationalized a large-scale survey study.

\subsubsection{Design of the Online Survey}
\label{sec:survey_method_design}

In designing the survey study, we adhered to the guidelines outlined by Dillman et al.~\cite{dillman2014internet} and Kitchenham et al.~\cite{kitchenham2008_PersonalOpinionSurveys}. Dillman et al.~\cite{dillman2014internet} tailored survey design method offers a structured framework for enhancing response rates and data quality. This method emphasizes principles such as customized survey distribution, clear questions, and active respondent engagement. Additionally, Kitchenham et al.~\cite{kitchenham2008_PersonalOpinionSurveys} empirical research guidelines in software engineering, particularly within the context of surveys, guided our choices regarding questionnaire design, sampling strategies, and data validation. By following these approaches, we aimed to reduce potential bias in participants' responses and ensure that the questions were easy to understand. Moreover, following ethical standards, we provided participants with an informed consent document and deliberately avoided collecting personal information, such as gender, age, or email addresses. Our focus was solely on gathering role and company-related information. Notably, the survey incorporated attention checks to identify participants who provided careless responses, thus for reliability reasons. 

Practically, we designed two questionnaires; the first (pre-screening survey) served as a preliminary screening tool to identify individuals matching our ideal criteria (better discussed in the following sections), while the second (main survey) was our survey. An iterative pilot test approach was employed to (1) evaluate the clarity and quality of the survey and (2) estimate the completion time. The sample group consisted of four individuals with at least one year of experience in the ML working context, and they were different from the individuals in the first phase, but shared similar characteristics. The ML practitioners validated the surveys and provided constructive feedback from various perspectives, which was incorporated to improve the overall quality of the questionnaires before they were distributed online.\footnote{More specifically, some redundant questions were removed, and typos that were identified were corrected.} Additionally, we analyzed the time it took pilot testers to complete the surveys, allowing us to estimate a completion time of 10 minutes.

\subsubsection{Structure of the Online Survey}

In terms of structure, the pre-screening survey comprised eight closed-ended questions designed to assess respondents' knowledge and experience regarding the sustainability of ML-enabled systems. Specifically, questions were included to determine their academic background in sustainability and ML. Examples of these questions include: \say{How familiar are you with the sustainability considerations of machine/deep learning?} and \say{Have you been involved in any projects where sustainability was a significant factor?}. Participants with limited knowledge of sustainability and ML were not selected for the main survey. This pre-screening process ensured that responses were gathered from individuals with relevant expertise.

Regarding the main survey, it was structured into five sections, aligning with the research objectives and mirroring the structure of the qualitative study. A summary of each section is provided in the following subsection.\footnote{For clarity, the detailed survey structure, including the full list of questions, is available in the online appendix \cite{appendix}.}

\begin{enumerate}
    \item \textbf{Background Information:} The first section collected demographic and professional background information, including the size of the participants' organizations, their education, years of experience, and the types of business systems they have developed. These questions provided an overview of their involvement in the development of ML-enabled systems.

    \smallskip
    \item \textbf{Description and Importance of Sustainability:} The second section, mapped to the first research question, explored participants' understanding, perception, and the perceived importance of sustainability across its various dimensions.

    \smallskip
    \item \textbf{Sustainability Approaches:} The third section, aligned with the second research question, focused on technical aspects. Participants were asked to evaluate the sustainability approaches identified in the interview study, rating their adoption frequency on a five-point Likert scale ranging from \textsl{`Never'} to \textsl{`Very Often'}. Additionally, they had the opportunity to report other sustainability approaches through a follow-up free-text box. Similar to the interview phase, participants were also asked to report tools and metrics used in conjunction with these approaches to support sustainability.

    \smallskip
    \item \textbf{Challenges in Sustainable ML Development:} The fourth section, corresponding to the third research question, addressed the challenges faced in ensuring the sustainability of ML systems. Participants rated the relevance of challenges identified through interviews on a five-point Likert scale ranging from \textsl{`Not Relevant'} to \textsl{`Very Relevant'}. They were also given the opportunity to describe additional challenges through a free-text response.

    \smallskip
    \item \textbf{Additional Insights:} In the final section, participants were provided with an open-ended text box to share any additional opinions or insights that could further inform the understanding of sustainability in practice. The survey concluded with a note of appreciation for their time and participation.
\end{enumerate}

\subsubsection{Survey Sample and Procedure}
\begin{figure}
    \centering
    \includegraphics[width=1\linewidth]{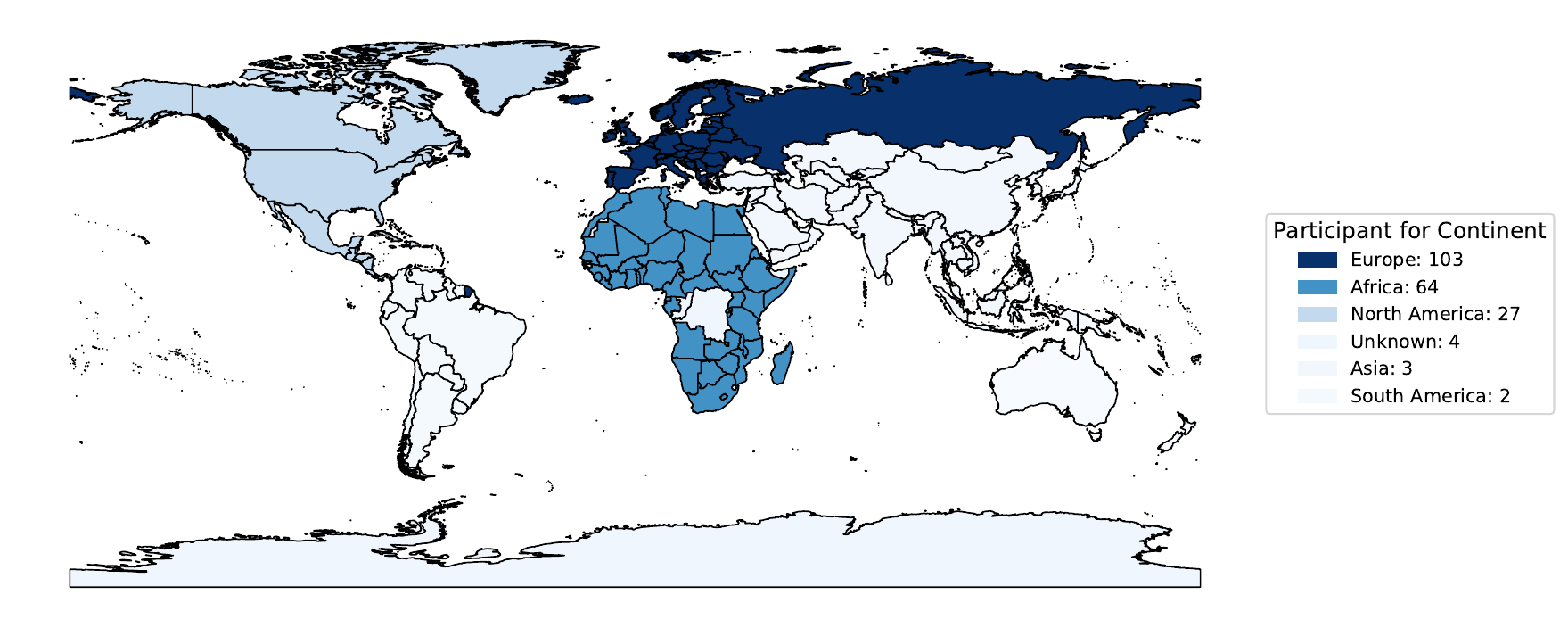}
    \caption{Survey participants distribution by continent.}
    \label{fig:continent}
\end{figure}

We selected \textsc{Prolific}\footnote{\textsc{Prolific}: \url{https://www.prolific.co/}} as the platform enabling the recruitment of participants and the collection of answers; the platform allows researchers to stipulate specific criteria for participants, i.e., in our case, it allowed limiting participation to ML practitioners.  In particular, this is a web-based research tool that has gained wide acceptance among researchers~\cite{reid2022_prolific_recommendations, ebert2022_prolific_recommendations} and is recognized as a high-quality recruitment platform when specific criteria—like those we adhered to as outlined—are met~\cite{eyal2021data}. It implements an opt-in strategy \cite{hunt2013participant}, which involves participants willingly agreeing to partake in the study. The use of \textsc{Prolific} influenced the design of the study; as such, we followed a set of recommendations provided in the literature~\cite{ebert2022_prolific_recommendations, ebert2022_prolific_recommendations, alami2024you}:

\begin{itemize}
    \item A repeated pre-screening method was used to enhance participant selection.
    \item The survey incorporated validation questions to verify that the respondents satisfied our selection criteria.
    \item A layered payment plan was employed, offering a \pounds 8 (“good”) reward for pre-screening and a \pounds 10 (“great”) reward for completing the main survey.\footnote{Prolific rates compensation in four levels—low, fair, good, and great—according to the payments made to participants.}
    \item As previously stated, questions designed to check attention were embedded within the questionnaires to verify the reliability of participants.
    \item We used a tool to detect the presence of AI-generated responses (specifically for open-ended questions) to identify any inappropriate behavior among participants.\footnote{The tool was ZeroGPT: \url{https://www.zerogpt.com} (February 2025)}
\end{itemize}

We employed a criterion-based purposive sampling approach for participant selection~\cite{palinkas2015purposeful}. Participants were chosen based on specific criteria related to their expertise and interest in ML systems. Those who indicated 'not interested', 'never worked with ML,' or 'demonstrated low interest and low knowledge of ML and sustainability', as per their responses in the pre-screening questionnaire, were excluded from the study. 
Moreover, \textsc{Prolific} provides historical metadata for each participant, such as the number of completed surveys and their approval rate. To guarantee data reliability, we restricted participation to individuals who had completed at least 20 previous tasks and maintained an approval rate of 90\% or higher—this was a proxy for ensuring worker quality and engagement. This sampling strategy increased our confidence in having selected individuals with relevant experience, credibility, and genuine interest in the subject matter.

On August 25, 2023, the pre-screening phase began with \textbf{300} participants recruited through \textsc{Prolific}. During this stage, five participants did not complete the survey and 44 were removed for failing attention check questions; to maintain the required sample size, \textsc{Prolific} provided 49 substitutes. In addition, 150 participants were excluded from the initial pool due to a lack of experience with ML systems and sustainability knowledge. Two days later, on August 27, 2023, the main survey was submitted to the remaining \textbf{150} participants. Of these, six participants were discarded after failing an attention check question, and 10 did not complete the questionnaire, resulting in a final set of \textbf{134} valid participants.


The initial data collection involved 134 participants, of which 64 ($\approx$48\%) were from Africa, with the remaining participants coming from various other regions. To incorporate a broader range of perspectives and address potential geographic and cultural confounding factors, we decided to repeat the recruitment process. In this second phase, we conducted another pre-screening and full survey, leveraging \textsc{Prolific}'s settings to target underrepresented regions and achieve a more balanced distribution of responses.
The second pre-screening was conducted on January 15, 2025, and 130 people were recruited. Of these, 31 did not complete the survey, and 51 were eliminated due to failed attention check questions or lack of experience in ML systems/sustainability knowledge (through \textsc{Prolific}, we got 43 substitutes). The second main survey was submitted on January 17, 2025 (two days later) on a sample of \textbf{91} people, of whom 17 did not complete the survey and 5 were discarded because they failed an attention check question.

As a consequence, the final set is composed of \textbf{203} valid participants. Figure \ref{fig:continent} illustrates the continent of each participant; however, four participants chose not to answer. Most of them have a Bachelor's degree (102), followed by a Master's degree (58), a High school diploma or equivalent (34), and a Ph.D. (9). Of the 203 participants, 39.41\% define themselves as Software Engineers, 16.26\% Data Scientists, 15.76\% Machine Learning Engineers, 15.27\% Data Engineers, and the remainder held other roles, \eg ML model Tester and Artificial Intelligence Engineer. Most work in Technical Domains (137) and the Economic and Political domains (60) while the remainder are in Socio-Cultural (49) and Medical and Scientific (32). The distribution of employees across company sizes is as follows: 10.34\% (21) of participants work in companies with fewer than 10 employees; 18.72\% (38) of participants are part of companies with a range of 11-50 employees; 28.08\% (57) of participants are employed in companies between 51 and 250 employees; 11.33\% (23) of participants are employed in organizations between 251 and 500 employees; and, finally, 31.03\% (63) of participants are employed in large companies with over 500 employees. Only one (0.49\%) participant does not remember the company size. The experience of developers is distributed as follows: 40.39\% (82) of developers have less than 2 years of experience, followed by 35.47\% (72) with 3-5 years of experience. A smaller number, 10.34\% (21), have 6 to 9 years of professional experience. Finally, those with more than 10 years' experience make up 13.79\% (28) of the sample. This distribution clarifies the various levels of experience within the group of surveyed developers.

\subsubsection{Data Cleaning and Analysis}

Upon reception of the participants' answers, the first author of the paper went through each of the individual responses to verify their reliability and credibility (using attention check questions, AI detector tools, and manual checking). The goal was to identify potential cases where participants may not have taken the task seriously or lacked sufficient experience to provide meaningful insights. From this analysis, 51 answers were found to be suspicious: therefore, a further discussion with the second and third authors of the paper was opened to decide whether these answers should have been accepted. At the end of the discussion, 14 of them were still considered valid. 

Once the data cleaning phase was completed, we analyzed the valid answers in two different manners. When considering close-ended questions, we computed statistics, e.g., frequency analysis and plots, to describe the trends observed in the participants' responses. As for open-ended questions, we analyzed them using \textit{open coding} and studied word frequencies to find patterns. The open coding exercise was conducted by the first three authors of the paper, who jointly collaborated toward the elicitation of the main concepts the answers let emerge. As in the first study, the process involved collaborative discussion and consensus to extract new concepts. Following the mixed-method approach, when a concept overlapped with one previously identified in the qualitative study, it was not newly created, but rather used to strengthen the existing concept.

\subsection{Connecting the Dots Between Qualitative and Quantitative Findings}
\label{sec:connectingDots}

The final phase of the study involved integrating and synthesizing the findings from both the qualitative and quantitative research stages. The survey results were systematically combined with the insights gathered from the semi-structured interviews using a qualitative synthesis approach. This phase aimed to align and cross-validate the findings, ensuring that the conclusions were robust, well-substantiated, and reflective of the diverse perspectives within the ML practitioner community. The iterative and collaborative integration process, whose results are detailed in Section \ref{sec:results}, began with a systematic comparison of key themes and patterns that emerged from the qualitative interviews against the statistical trends identified in the survey responses. Areas of convergence were identified, confirming aspects of sustainability practices that were consistently highlighted across both data sources. For instance, the main challenges identified in the interviews were validated by the prevalence of these concerns in the survey results. Conversely, areas of divergence were critically examined to understand discrepancies. Instances where survey participants reported different perspectives from those expressed in the interviews were analyzed through follow-up discussions among the authors. Such cases were explored in depth, considering contextual factors such as differences in professional roles, company size, and experience levels. 

Following this synthesis, the integrated results were further contextualized within the existing body of literature (Section \ref{sec:discussion}). A comparative analysis was conducted between the findings of this study and prior related works on sustainability in ML-enabled systems.
In particular, a triangulation process was carried out to assess the plausibility of the approaches and challenges reported by ML practitioners. Specifically, for each observation emerging from interviews and surveys, we reviewed the existing literature to identify supporting evidence or comparable results. When relevant studies were identified, e.g., confirming the sustainability benefits of cloud adoption, model compression, or use of pre-trained models, they were cited directly in the results section. Conversely, in cases where no academic evidence was available, we explicitly acknowledged this absence and framed the practitioner insights as exploratory, suggesting them as potential directions for future empirical research. This process allows readers to clearly distinguish between practices grounded in prior literature and those primarily based on field observations, thereby enhancing both transparency and interpretive depth. The process described above aligned with the principles of Grounded Theory~\cite{9520216, strauss1997_grounded_theory, strauss1990_basics_of_qualitative_research, glaser2017_original_grounded_theory}, ensuring that the study’s contributions extended beyond mere descriptive insights and instead provided an enriched understanding of sustainability in ML practices. The literature comparison helped position the findings within the broader discourse, identifying how our study corroborated previous findings, other than areas where this study advanced the state of knowledge in a non-biased manner.


\section{Analysis of the results}
\label{sec:results}

In this section, we report the results of our analysis on both the interviews and the survey.

\subsection{RQ$_1$ — Sustainability Perception}

The first research question aimed to investigate how ML practitioners perceive sustainability and how they prioritize its different dimensions in terms of importance. To mitigate potential bias in participants' responses, such as social desirability or acquiescence, we adopted a multi-step questioning strategy. First, participants were asked to describe sustainability in their own words, without predefined categories. Only after this, we presented structured questions, including Likert-scale items and a ranking question, to capture both perceived importance and forced prioritization of sustainability dimensions.

From the interviews, P3, P4, P5, and P6 described sustainability in ML software as a trade-off, emphasizing that achieving sustainability often requires a slight reduction in efficiency to minimize negative environmental and societal impacts. Additionally, P3 and P6 highlighted that “good” sustainability entails \textit{\say{making processes efficient while maintaining the same level of effectiveness}}.

Conversely, P1 and P7 perceived sustainability less as a trade-off and more as an ethical responsibility. For instance, P1 defined sustainability as \textit{\say{something that causes as little harm as possible to society and the environment}}, while P7 described it as \textit{\say{the awareness and understanding of the potential consequences that an action may have in the future}}. Unlike the trade-off perspective, these views underscore the moral obligation to integrate sustainability into decision-making, emphasizing the responsibility of those involved in the process. Furthermore, when asked about the relative importance of sustainability dimensions (\ie social, environmental, and economic), participants consistently ranked the economic dimension as the least important, followed by the social dimension, with the environmental dimension being considered the most critical.

\begin{figure}
     \centering
    \includegraphics[width=1\linewidth]{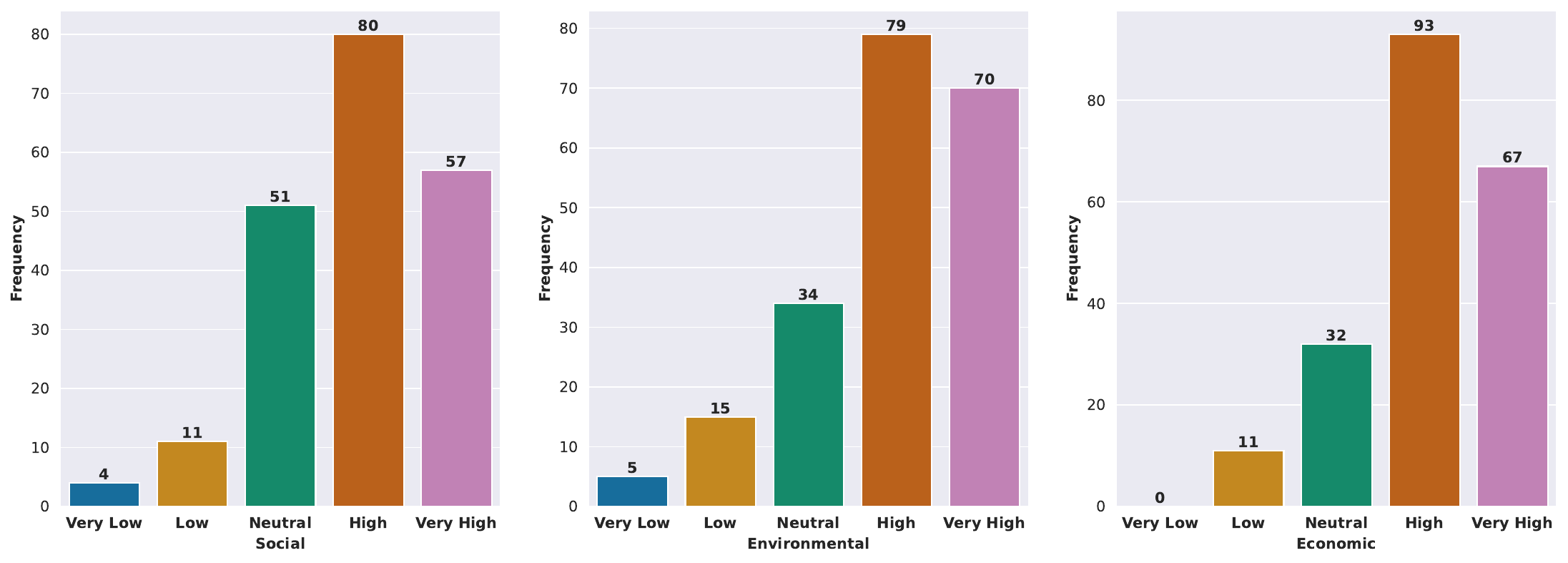}
    \caption{Sustainability rate for survey respondents.}
    \label{fig:respondents_rate}
    \end{figure}
    
    \begin{figure}
    \centering
    \includegraphics[width=1\linewidth]{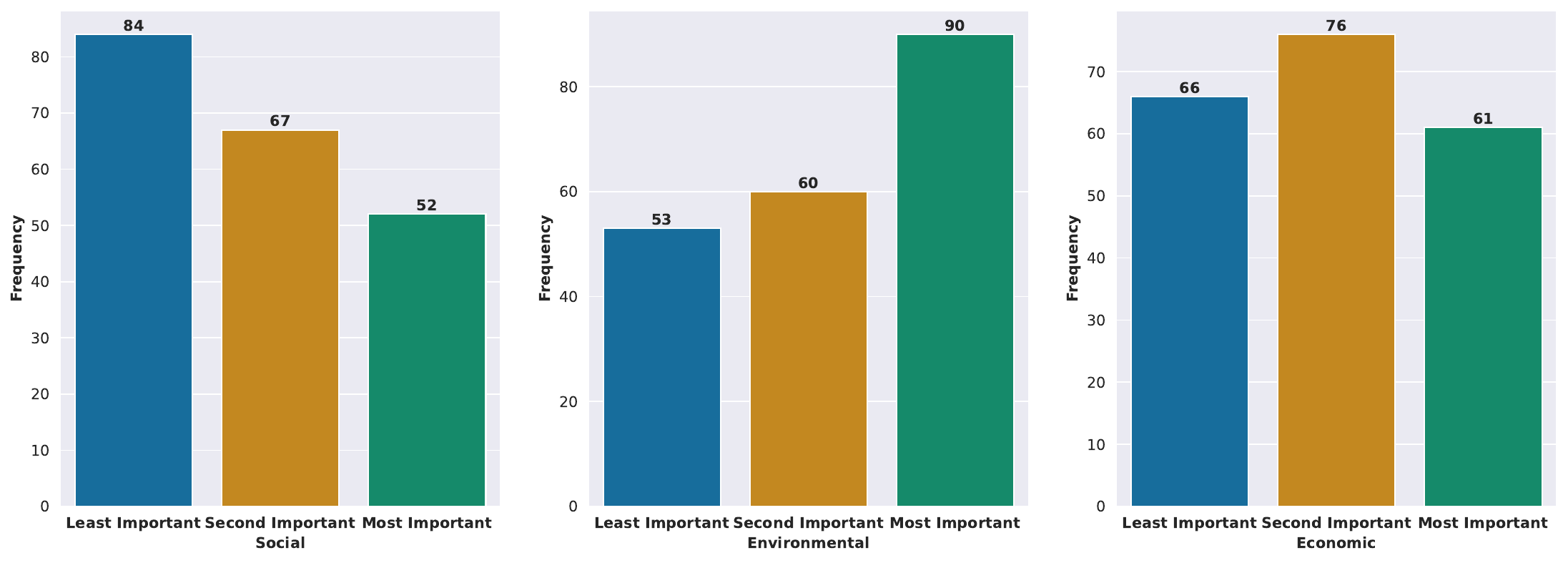}
    \caption{Sustainability importance for survey respondents.}
    \label{fig:respondents_imp}
\end{figure}

\begin{figure}
\centering
\includegraphics[width=0.5\linewidth]{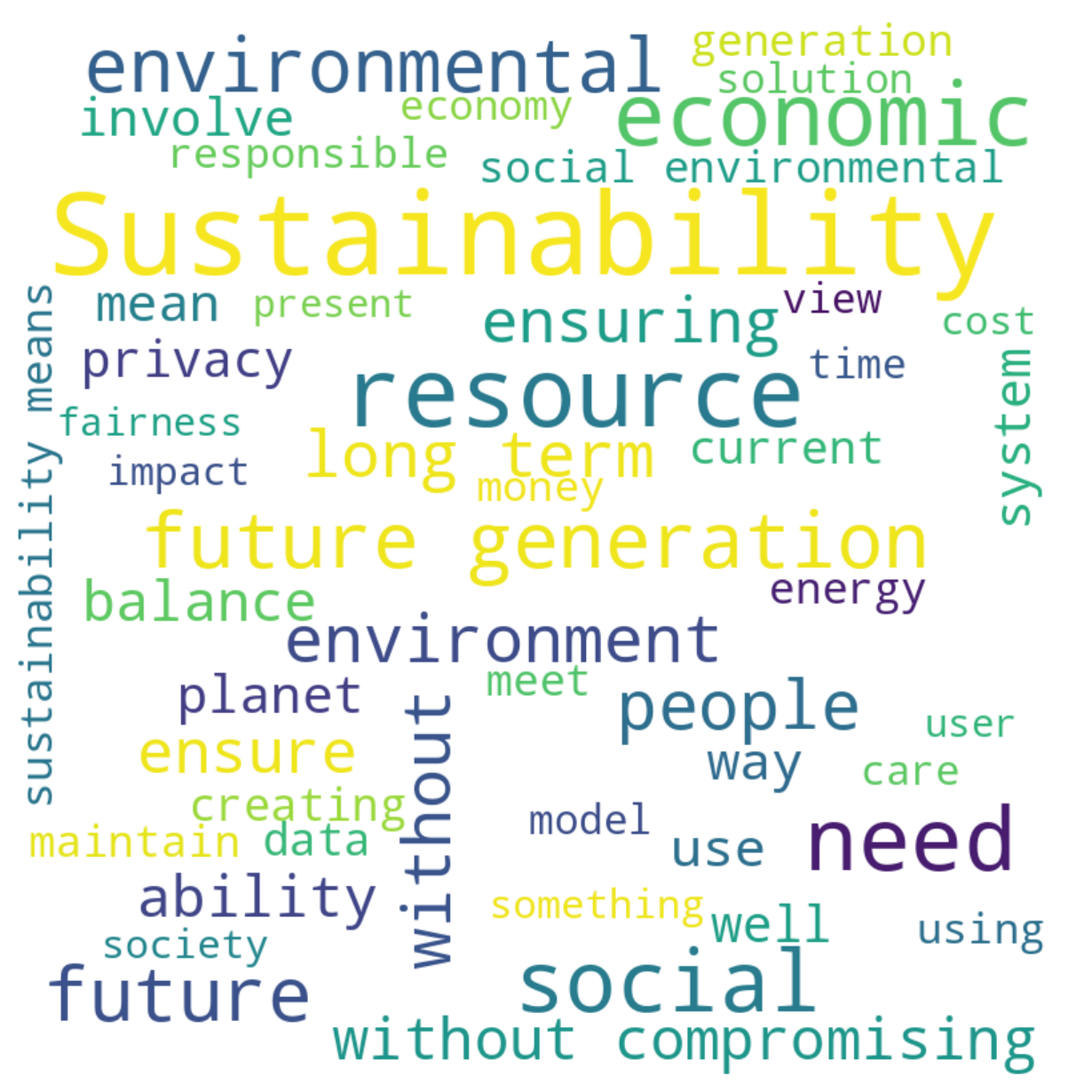}
\caption{Word cloud generating analyzing survey participants responses for the first research question.}
\label{fig:cloud}
\end{figure}

Moving to the survey, the word cloud illustrated in Figure \ref{fig:cloud}, was crafted to provide a concise and insightful representation of the predominant terms encapsulating the 184 respondents' perceptions of sustainability. We have extracted the most frequently used terms from each respondent's definition and calculated their respective frequencies. The top three terms, along with their frequencies, are \emph{Future/Future generations} (34.62\%), \emph{Social} (27.07\%), and \emph{Resources} (23.33\%).
Figures \ref{fig:respondents_rate} and \ref{fig:respondents_imp} display the rates and the importance assigned to each sustainability dimension (i.e., social, environmental, and economic) by the survey respondents. As depicted in Figure~\ref{fig:respondents_rate}, practitioners predominantly perceive all three dimensions as highly or very highly important. In terms of ranking these dimensions, Figure~\ref{fig:respondents_imp} reveals that the environmental dimension is considered the most important in the majority of cases (90 out of 203 respondents), followed by the economic dimension, with the social dimension being rated as the least important in 52 cases. 

Overall, the findings indicate that ML practitioners in our study perceive sustainability through two primary lenses: as a trade-off involving efficiency reductions to mitigate environmental and societal impact, or as an ethical responsibility emphasizing long-term consequences and minimal harm. While interviewees provided nuanced perspectives on these aspects, survey results further highlighted the prominence of sustainability from practitioners’ viewpoints. The word cloud analysis (Figure~\ref{fig:cloud}) revealed that terms such as \emph{Future/Future generations}, \emph{Social}, and \emph{Resources} were most frequently mentioned, reinforcing the forward-looking and multidimensional nature of sustainability concerns. Additionally, quantitative findings (Figures~\ref{fig:respondents_rate} and~\ref{fig:respondents_imp}) demonstrated that practitioners consider all three dimensions—social, environmental, and economic—as important, albeit with varying degrees of emphasis. In our study, while the environmental dimension was ranked as the most critical by the majority, the economic and social dimensions were perceived as secondary, reflecting a broad yet stratified prioritization of sustainability concerns in the ML approach.

\definitionbox{RQ$_1$ — Sustainability Perception}{In our study, ML practitioners perceive sustainability either as a trade-off requiring efficiency reductions or as an ethical responsibility focused on long-term impact. Survey results reinforce this dual view, with key terms like \emph{Future/Future generations}, \emph{Social}, and \emph{Resources} frequently mentioned. While all three sustainability dimensions are considered important, the environmental dimension is ranked highest, followed by the economic and social aspects.}



\begin{table}[h]
\centering
\footnotesize
\caption{Categorization of Sustainable  Approaches}
\label{tab:approaches}
\resizebox{\textwidth}{!}{%
\rowcolors{1}{gray!15}{white}
\begin{tabular}{|p{3cm}|p{3cm}|p{8cm}|}
\rowcolor{black}
\hline
\textcolor{white}{\textbf{Category}} & \textcolor{white}{\textbf{Description}} & \textcolor{white}{\textbf{Approaches}} \\\hline
Cost and Resource Efficiency & Approaches targeting reductions in financial cost or resource consumption of ML-enabled systems. 
& (A.1) Selecting Efficient Libraries and Languages \\
\cline{3-3}
& & (A.2) Optimizing Code to Minimize Resources \\
\cline{3-3}
& & (A.3) Adopt Strategies for Efficient Management of ML Costs \\
\hline
Data Management & Approaches focused on improving data quality and efficiency in the preparation phase of ML-enabled systems.
& (A.4) Consider Ethical Data Management \\
\cline{3-3}
& & (A.5) Clean and Optimize Dataset \\
\cline{3-3}
& & (A.6) Consider Train with Synthetic Data \\
\hline

Model Optimization & Approaches that aim to reduce the computational complexity and resource demands of ML-enabled systems. 
& (A.7) Consider Adopting Pre-trained Models \\
\cline{3-3}
& & (A.8) Consider Streamlining Models for Efficiency and Cost Reduction \\
\cline{3-3}
& & (A.9) Preferring Simplicity in System Design \\
\cline{3-3}
& & (A.10) Consider Delegating Learning to Exploration \\
\cline{3-3}
& & (A.17) Consider Knowledge Distillation \\
\hline

Hardware and Infrastructure & Approaches that improve sustainability by adapting or optimizing the underlying computational environment of ML-enabled systems. 
& (A.11) Consider Adopting Cloud Technologies \\
\cline{3-3}
& & (A.12) Consider Implementing Offline Predictions \\
\cline{3-3}
& & (A.13) Evaluating the Suitable Model for the Task \\
\cline{3-3}
& & (A.14) Operational Sustainability in AI Devices \\
\cline{3-3}
& & (A.18) Consider New Hardware \\
\hline
Transparency and Monitoring & Approaches aimed at increasing the observability and ethical accountability of ML-enabled systems. 
& (A.15) Consider Measuring Ethical Aspects \\
\cline{3-3}
& & (A.16) Track Sustainability Metrics \\
\cline{3-3}
& & (A.19) Consider Integrate Model Explainability \\
\hline
\end{tabular}
}
\end{table}

\subsection{RQ$_2$ — Sustainability Approaches}
The second research question aimed to explore the approaches used to support sustainability in software development, with a particular focus on the tools and metrics adopted by ML practitioners. To further improve the clarity and interpretability of our findings for \textbf{RQ$_2$}, we grouped the identified sustainability-related approaches into five thematic categories. This categorization was developed through a bottom-up analysis of the approaches’ functional goals and their application context. The five resulting categories are: (i)\textsl{`Cost and Resource Efficiency'}, (ii) \textsl{`Data Management'}, (iii) \textsl{`Model Optimization'}, (iv) \textsl{`Hardware and Infrastructure'}, and (v) \textsl{`Transparency and Monitoring'}. Table \ref{tab:approaches} summarizes the groups with short descriptions and the list of associated approaches. It is worth remarking that the table reports all the approaches emerged in response to \textbf{RQ$_2$}, including those emerging from both interviews and the survey. In Section \ref{sec:interviewResults_Approaches}, we focus on the 16 approaches derived from the interviews (\textbf{A.1} to \textbf{A.16}). The remaining three approaches (\textbf{A.17} to \textbf{A.19}), which emerged directly from survey responses, are discussed separately in Section~\ref{sec:surveyResults_Approaches} to explicitly preserve the distinction between data sources. Each approach is accompanied by a list of references when supported by scientific literature, as identified through our triangulation process described in Section~\ref{sec:connectingDots}. When no supporting literature was found, we explicitly acknowledge the lack of evidence and frame the practice as exploratory.

\subsubsection{Approaches from the Interviews}
\label{sec:interviewResults_Approaches}
From our interviews, participants described several approaches (A) that influence sustainability. To enrich the discussion of these results, we include selected quotations that illustrate the participants’ original phrasing and provide contextual grounding for each approach. 

\paragraph{\textbf{\underline{Cost and Resource Efficiency}}} This group comprises three approaches that focus on optimizing code, minimizing unnecessary computation, and selecting efficient execution strategies to lower energy use and financial costs.

\begin{description}[leftmargin=0.3cm]
    
    \smallskip
    \item \faLeaf\ \textbf{(A.1) Selecting Efficient Libraries and Languages \cite{järvenpää2023synthesis,marini2024green}:} 
    P2, P3, and P7 report the approach to reducing power consumption and costs by carefully selecting efficient libraries and languages. For instance, P1, when analyzing the execution time, reported \textit{\say{From an economic and environmental point of view, Python is the worst.}}. Additionally, P1 described \textit{\say{Is very simple and full of libraries to use for convenience in the first phase without standing there and wasting too much time on it. For the training part, we use C language.}}. Practitioners continually seek ways to reduce resource consumption through model optimization or efficient training techniques.
    
    \smallskip
    \item \faLeaf\ \textbf{(A.2) Optimizing Code to Minimize Resources \cite{chiang2023dynamic}:}
    P1, P2, P3, P5, and P6 minimize expenses and energy usage through strategies such as reducing hardware utilization and preventing the wastage of unused resources. They achieve this by optimizing the quality of their code for both CPU and GPU usage in the case of P1 and P2. P1 highlighted \textit{\say{It is necessary to optimize the use of the internal memory of the graphics processing units (GPUs), as the amount available is often insufficient.}} Additionally, says \textit{\say{While GPUs are designed to optimize tensor processing, images are processed by the CPU, creating a performance problem and limitations.}}. 

    \smallskip
    \item \faLeaf\ \textbf{(A.3) Adopt Strategies for Efficient Management of ML Costs \cite{gonzalez2025impact,reguero2025energy}:} All interviewees share a common objective of minimizing energy consumption and reducing training time, indirectly leading to decreased development costs. These approaches are consistently implemented across the board, with some participants like P1 and P4 halting training once a specific accuracy threshold is reached, thereby respecting business constraints and emphasizing the importance of improving data quality rather than continuing to train the model. Additionally, P6 reported that, \textit{\say{for example, the drones we know fly for 1:45 hours with a 15-watt decision system. If the decision system is at 10 watts, the flight time can increase from 1:45 to 1:52 hours. It means they cover a larger area, so not only do you create a solution that has a smaller impact from an energy standpoint but also a smaller impact on the number of drones to be released because if these fly for 1:45 hours, I need three, maybe if they fly for 2 hours, I can only use two.}}

\end{description}

\paragraph{\textbf{\underline{Data Management}}} This group includes three approaches centered on improving the sustainability of ML systems through better data handling. The practices focus on data preparation, quality assurance, and preprocessing.

\begin{description}[leftmargin=0.3cm]

    \smallskip
    \item \faLeaf\ \textbf{(A.4) Consider Ethical Data Management \cite{voria2025fairness,xu2021privacy}:} P5 highlighted the importance of anonymizing sensitive user data before training ML models, using techniques such as MD5 hashing and Hashbytes. This approach is aimed at promoting social sustainability by safeguarding user privacy. P5 further explained:  \textit{\say{The data that we then provide to the business is anonymized, hiding personal information either through encryption or through dummies.}}

    \smallskip
    \item \faLeaf\ \textbf{(A.5) Clean and Optimize Dataset \cite{järvenpää2023synthesis,salehi2023data}:} P1, P2, P5, and P7 emphasized that before training an ML model, there is a need for high-quality data and the need to reduce the size of the dataset. This process is done through the removal of redundant values that increase the use of energy and storage resources to train the models.

    \smallskip
    \item \faLeaf\ \textbf{(A.6) Consider Train with Synthetic Data \cite{salehi2023data}:} Due to the scarcity of authentic data, P6 optimizes the development of their systems in terms of both time and cost efficiency by employing synthetic data for training purposes. Once a model achieves a specified level of accuracy, it is deployed into a physical device. After the model is deployed into the system, new real-world data is collected, which is then used for further refinement and improvement of the system through continuous training. This strategy accelerates the initial model-building process and facilitates continuous model performance improvement over time.

\end{description}

\paragraph{\textbf{\underline{Model Optimization}}} The interview participants mentioned four approaches aimed at improving sustainability by optimizing ML models. These practices involve selecting simpler architectures, reducing model size through compression or pruning, and balancing performance with resource efficiency to limit computational and environmental costs.

\begin{description}[leftmargin=0.3cm]

    \smallskip
    \item \faLeaf\ \textbf{(A.7) Consider Adopting Pre-trained Models \cite{järvenpää2023synthesis,verdecchia2023systematic}:}
    P1, P2, P6, P7 and P8 prioritize using pre-trained models as a cost-effective strategy, which reduces development costs and lowers energy consumption than training from scratch. 

    \smallskip
    \item \faLeaf\ \textbf{(A.8) Consider Streamlining Models for Efficiency and Cost Reduction \cite{järvenpää2023synthesis,verdecchia2023systematic}:} Sustainability strategies, as mentioned by P3, encourage critical evaluation of patterns. This involves identifying “lazy” neurons in neural networks or unnecessary parts of models that can be eliminated. This approach streamlines models, enhancing efficiency without compromising performance. On this, P3 highlights: \textit{\say{When we take state-of-the-art open source models, especially academics, tend to have less attention, and maybe the network works, but maybe there are a lot of lazy and useless neurons like that, maybe after the network loading is a little slower.}}.  Additionally, P6 highlights using quantization and low-rank adapter to reduce the model size, describing \textit{\say{Good functions that filter the input also go a long way to reducing power consumption by ensuring the same performance as the model grows. However, this thing we have seen fails because the number of the parameters is the real bottleneck more than the input size}} 

    \smallskip
    \item \faLeaf\ \textbf{(A.9) Preferring Simplicity in System Design:} Participants P1, P2, P3, and P6 opt for simpler model architectures, recognizing that they are cost-effective and easier to maintain while still delivering effective results. P3 reported \textit{\say{I always try to think in a simpler system. A simple system consumes fewer resources, which has all the good impact.}}. Complexity can lead to resource wastage and challenges in long-term management and maintenance. Sustainability strategies encourage keeping systems simple, reducing resource consumption, and facilitating ongoing maintenance. Our triangulation process revealed that this approach has not yet been examined in the scientific literature - hence, it can be considered an emerging practice.

    \smallskip
    \item \faLeaf\ \textbf{(A.10) Consider Delegating Learning to Exploration \cite{kim2020autoscale,järvenpää2023synthesis}:} In situations with limited data, participant P3 incorporates reinforcement learning and employ lighter models to reduce cost; P3 said \textit{\say{If it is possible to use a reinforcement learning approach, we prefer it. In general, we delegate a lot of the learning work to the exploration part, which is always more efficient than the training part, which remains inefficient because, usually, the networks are very small in reinforcement learning models.}}.

    \end{description}

\paragraph{\textbf{\underline{Hardware and Infrastructure}}} This group comprises four approaches focused on enhancing sustainability by acting on the underlying computational environment. These practices involve selecting energy-efficient hardware, optimizing infrastructure configurations, and leveraging cloud or edge resources to reduce resource waste and improve overall system efficiency.

  \begin{description}[leftmargin=0.3cm]  
    \item \faLeaf\ \textbf{(A.11) Consider Adopting Cloud Technologies \cite{järvenpää2023synthesis,verdecchia2023systematic}:} to reduce energy consumption and minimize development costs, participants P2, P3, P4, and P8 rely on the usage of cloud technologies. By doing so, practitioners could benefit from machines optimized for parallel computing and minimize environmental impact. Participant P1 reported \textit{\say{Currently, we use an in-house machine, which requires a server room with a constant low temperature to handle the heat generated. The associated costs are significant in purchasing and maintaining the machine. I recognize that using cloud computing can be beneficial in terms of cost and energy efficiency. Resources in the cloud are used more efficiently than on a dedicated machine, reducing waste and optimizing overall energy consumption}}. 

    \smallskip
    \item \faLeaf\ \textbf{(A.12) Consider Implementing Offline Predictions:} 
    An alternative approach suggested by P1 to reduce costs and energy consumption involves implementing offline predictions on customer PCs. Furthermore, both P1 and P4 prioritize allocating machine resources based on task importance as another strategy for efficiency. According to our triangulation process, this approach has not yet been addressed in the scientific literature. It may therefore be seen as an emerging approach that warrants further investigation by the research community.
    
    \smallskip        
    \item \faLeaf\ \textbf{(A.13) Evaluating the Suitable Model for the Task \cite{tabbakh2024towards}:} Before embarking on solution development, participants P1, P2, P6, P7, and P8 meticulously evaluate and select the most appropriate model for a specific task. Their focus is on cost reduction and performance enhancement. 
    P4 reports \textit{\say{If I have a model of 8 GB and another one of 4 GB. Even if I lose two percentage points, whatever, I take the smaller one, so I don't have to get the bigger machine.}}. P1 described for the preprocessing phase \textit{\say{Optimization is a key element in managing time, memory, and cost in the machine learning process, with particular complexity in choosing the architecture and optimizing the processing.}}.
    As emphasized by P1, P2, and P3, sustainability strategies drive the adoption of more resource-efficient development approaches. This is particularly crucial in machine learning, where training models demand substantial time and computing power. Practitioners are motivated to find ways to reduce training time and optimize hardware utilization.
    
    \smallskip
    \item \faLeaf\ \textbf{(A.14) Operational Sustainability in AI Devices \cite{chiang2023dynamic,castanyer2024design}:} P6 highlighted the need to operationalize sustainability in IoT or memory-constrained systems, where input and model sizes are limited. In particular, P6 highlight \textit{if you load a model that goes to memory saturation or the number of inferences required, it will shut down.} This approach improves the use of limited memory and processing capacity and reduces power consumption within constrained systems.

    \end{description}
    

\paragraph{\textbf{\underline{Transparency and Monitoring}}} This group includes two approaches that aim to enhance the sustainability of ML systems by improving their observability and ethical accountability.

  \begin{description}[leftmargin=0.3cm]  
    \smallskip
    \item \faLeaf\ \textbf{(A.15) Consider Measuring Ethical Aspects \cite{ferrara2024fairness,voria2025fairness}:} Moreover, participant P3 takes steps to measure equity metrics with the objective of addressing social discrimination and fostering responsible approaches in their endeavors.

    \smallskip
    \item \faLeaf\ \textbf{(A.16) Track Sustainability Metrics \cite{järvenpää2023synthesis,verdecchia2023systematic}:} 
    Interviewees emphasized the importance of integrating sustainability metrics throughout the development lifecycle of ML-enabled systems. Monitoring these metrics offers valuable insights into resource consumption and potential system biases, enabling process optimization, cost savings, and the development/operation of more sustainable and responsible systems.

\end{description}

\subsubsection{Approaches from the Online Survey}
\label{sec:surveyResults_Approaches}
Moving to the questionnaire, participants were first asked to indicate whether they had implemented approaches that emerged in the interviews. 

The majority of respondents reported using techniques aimed at enhancing computational efficiency: 66.50\% (135) indicated that their teams optimize CPU/GPU usage to reduce energy consumption, while 56.65\% (115) reported utilizing cloud technologies instead of traditional on-premise solutions. Moreover, model optimization strategies were also widely adopted. Notably, 36.45\% (74) of participants focused on reducing model size, and 35.47\% (72) actively eliminated unnecessary information, such as underutilized "lazy" neurons. Additionally, 33.50\% (68) adopted simpler, lighter architectures, and 30.54\% (62) leveraged pre-trained models. Furthermore, 24.63\% (50) employed reinforcement learning techniques as part of their optimization efforts. As a final note, other approaches were implemented to a lesser extent. For instance, 25.12\% (51) of respondents measured and addressed fairness issues within their models, and 23.15\% (47) reduced training time once the desired accuracy was achieved. Strategies such as training with synthetic data 7.39\% (15), reducing dataset size 5.91\% (12), and improving the sustainability of devices that integrate AI components 5.91\% (12) were adopted even less frequently. Lastly, a small portion of the sample (3.45\%, 7) reported that they had not implemented any of the specified methodologies. Moreover, survey participants could also report other approaches. From the analysis of the survey responses, we elicited three additional approaches. These contributions complement the interview findings and map to three different groups in Table~\ref{tab:approaches}, namely: \textsl{`Model Optimization'} (\textbf{A.17}), \textsl{`Hardware and Infrastructure'} (\textbf{A.18}), and (v) \textsl{`Transparency and Monitoring'} (\textbf{A.19}). For all three approaches, our triangulation process identified relevant supporting literature, which we reference in the corresponding descriptions to substantiate their contribution to sustainable ML development.

\paragraph{\textbf{\underline{Model Optimization}}} This category now includes one additional approach.
\begin{description}[leftmargin=0.3cm]
    \item \faLeaf\ \textbf{(A.17) Consider Knowledge Distillation \cite{järvenpää2023synthesis,verdecchia2023systematic}:} Participants emphasize the need to design ML-enabled systems to balance accuracy and resource utilization. They highlighted the use of knowledge distillation to reduce model size without compromising performance. This approach could reduce model size and computational complexity during both training and inference. 
\end{description}

\paragraph{\textbf{\underline{Hardware and Infrastructure}}} This category now includes one additional approach.
\begin{description}[leftmargin=0.3cm]
    \item \faLeaf\ \textbf{(A.18) Consider New Hardware \cite{tabbakh2024towards}:} If it is not possible to adopt \textbf{(A.11)}, developers emphasize not to use old hardware for too long. Sometimes, it is worth spending money on more efficient technologies and processes to reduce development costs.

\end{description}
\paragraph{\textbf{\underline{Transparency and Monitoring}}} This category now includes one additional approach.
\begin{description}[leftmargin=0.3cm]
    \item \faLeaf\ \textbf{(A.19) Consider Integrate Model Explainability \cite{habiba2025ml}:} Developers have emphasized on creating interpretable models, balancing complexity and transparency. Simpler, interpretable architectures may be preferred over “black-box” approaches unless necessary. Additionally, tools such as SHAP and LIME can be used to assess the contribution of individual features to model predictions, uncovering potential biases and areas for improvement. Interpretability increases stakeholder confidence and better meet customer needs, as well as facilitates debugging and refinement of ML-enabled systems.
\end{description}

\begin{table}[h]
    \centering
    \caption{Frequency of sustainability tool adoption by survey participants.}
    \label{table:list_tools}
    \resizebox{\textwidth}{!}{%
    \rowcolors{1}{gray!15}{white}
    {
    \begin{tabular}{|l|c|c|} \hline 
    \rowcolor{black}
    \textcolor{white}{\textbf{Sustainability Tools}} & \textcolor{white}{\textbf{Frequency}} & \textcolor{white}{\textbf{Sustainability Dimension}} \\\hline
    \textit{Microsoft Fair Learn} &  \textbf{41.38\% (84)} & Social \\\hline
    \textit{CodeCarbon} &  \textbf{33.99\% (69)} & Environmental \\\hline
    \textit{IBM 360 degree toolkit} &  \textbf{25.62\% (52)} & Social \\\hline
    \textit{What-if Tool from Google} &  \textbf{23.65\% (48)} & Social \\\hline
    \textit{Never Used Tools} &  \textbf{22.17\% (45)} & -- \\\hline
    \textit{AI-Fairness 360} &  \textbf{20.69\% (42)} & Social \\\hline
    \textit{Running Average Power Limit (RAPL)} &  \textbf{12.81\% (26)} & Environmental \\\hline
    \textit{Aequitas} &  \textbf{6.90\% (14)} & Social \\\hline
    \textit{NVIDIA Triton Inference Server} &  \textbf{5.42\% (11)} & Environmental \\\hline
    \textit{SHapley Additive exPlanations (SHAP)} &  \textbf{0.98\% (2)} & Social \\\hline
    \textit{Local Interpretable Model-agnostic Explanations (LIME)} &  \textbf{0.49\% (1)} & Social \\\hline
    \textit{Internal Developed Software} &  \textbf{0.49\% (1)} & -- \\\hline
    \textit{Databricks} &  \textbf{0.49\% (1)} & Economic \\\hline
    \textit{Microsoft Sustainability Calculator} &  \textbf{0.49\% (1)} & Environmental \\\hline
    \end{tabular}}}
\end{table}

\subsubsection{Sustainability Metrics and Tools}

Other than discussing approaches, we also discussed and collected sustainability metrics and tool used by ML practitioners.

Regarding the metrics, most of the information was derived from the survey study. In the interviews, P2 and P3 mentioned measuring training time, while P6 referred to the use of \textsl{Triton tools}, a benchmarking suite designed to assess model performance (inference time) and energy consumption. In contrast, the survey participants identified a broader range of metrics, particularly within the environmental dimension, including energy consumption, CO\textsubscript{2} emissions, and water usage. Metrics related to the social dimension encompassed employee satisfaction and educational attainment rates. Within the economic dimension, commonly adopted metrics included return on investment (ROI), key performance indicators (KPIs), and cost-benefit analysis measures for the models under evaluation. A comprehensive report detailing all the identified metrics is available in the online supplementary material \cite{appendix}. It is important to note that ROI was not used to define the concept of economic sustainability in our study, but rather as a practical proxy frequently employed by practitioners to evaluate cost-effectiveness. While ROI captures short- or medium-term financial performance, it does not account for broader economic sustainability objectives, which emphasize long-term viability and resource efficiency.

Moving to the tools, most of the gathered information comes from the survey study, with only P6 (from the interview) mentioning the \textsc{Triton tools}. Table \ref{table:list_tools} provides an overview of the tool usage's frequency among ML practitioners (according to the survey participants) and the sustainability dimension they impact. The most commonly employed tool is \textsc{Microsoft Fair Learn}, followed by \textsc{CodeCarbon}, \textsc{Google's What-if Tool},  \textsc{SHAP}, \textsc{LIME}, \textsc{Databricks}, and \textsc{Microsoft Sustainability Calculator}. In contrast, 22.17\% (45) respondents reported never using any tools. Regarding frameworks no interview participants use framework, instead, for survey participants 41.87\% (85) never used frameworks. Among those who do, the most prevalent frameworks are \textsc{IFRS} (33.99\%), \textsc{CFD} (24.63\%), and \textsc{Gri standards} (19.70\%), with only 0.49\% of respondents relying on internally developed software and ED Framework (0.49\%).

\definitionbox{RQ$_2$ — Sustainability Approaches}{We identified 19 sustainable approaches, their effects on the development of ML-enabled systems, and categorized them into five groups. The most frequently used approaches are CPU/GPU optimization to reduce power consumption and the adoption of cloud technologies. We also quantified the frequency with which these sustainable approaches, tools, and frameworks are used. The result is a mapping of ML practitioners' current knowledge of sustainability in our study, providing insights into areas of research worth exploring in the future.}

\subsection{RQ$_3$ — Sustainability Challenges}
We then focused on the challenges encountered by ML practitioners in integrating sustainability into software development. We grouped the 14 identified challenges into four thematic categories through a bottom-up analysis of their content and application context. The  resulting categories are: (i) \textsl{`Norms, Standards, and Guidelines'}, (ii) \textsl{`Organizational and Economic Support'}, (iii) \textsl{`Skills and Human Resources'}, and (iv) \textsl{`Responsibility and Ethics'}. Table \ref{tab:challenges} summarizes these categories with brief descriptions and the list of associated challenges. Similarly to the organization of the results for \textbf{RQ$_2$}, the table reports all the challenges that emerged in response to \textbf{RQ$_3$}, including those from both interviews and the survey. In Section \ref{sec:interviewResults_Approaches_Challenges}, we focus on the 14 challenges derived from the interviews (\textbf{C.1} to \textbf{C.8}). The remaining six challenges (\textbf{A.9} to \textbf{A.14}), which emerged from the survey, are discussed separately in Section~\ref{sec:surveyResults_Approaches_Challenges}. Also in this case, each challenge is accompanied by a list of references when supported by scientific literature. When no supporting literature was found, we explicitly acknowledge this lack of evidence and frame the challenges as exploratory.

\subsubsection{Challenges from the Interviews}
\label{sec:interviewResults_Approaches_Challenges}
The interviews revealed challenges (C) that practitioners faced when dealing with sustainability.

\begin{table}[h]
\centering
\footnotesize
\caption{Categorization of Sustainable  Challenges}
\label{tab:challenges}
\rowcolors{1}{gray!15}{white}
\begin{tabular}{|p{3cm}|p{3.5cm}|p{6cm}|}
\hline
\rowcolor{black}
\textcolor{white}{\textbf{Category}} & \textcolor{white}{\textbf{Description}} & \textcolor{white}{\textbf{Challenges}} \\
\hline
Norms, Standards, and Guidelines & Lack of formal references and criteria to guide sustainable development & (C.1) Absence of Standardized Metrics \\
\cline{3-3}
& & (C.2) Limited Integration Method \\
\cline{3-3}
& & (C.9) Lack of Sustainability Standard \\
\cline{3-3}
& & (C.10) Sustainability Lacking in Design \\
\hline

Organizational and Economic Support & Missing support from institutions, funding, or client demand & (C.3) Lack of Awareness \\
\cline{3-3}
& & (C.4) Lack of Government Incentives \\
\cline{3-3}
& & (C.5) Lack of Customer Demand \\
\cline{3-3}
& & (C.11) Lack of Sufficient Financial Support \\
\hline

Skills and Human Resources & Limited knowledge, training, or frameworks for sustainability & (C.6) Lack of Software Frameworks \\
\cline{3-3}
& & (C.7) Lack of Educational Resources \\
\cline{3-3}
& & (C.12) Lack of Staff Training \\
\cline{3-3}
& & (C.13) Sustainability Slows Down Software Development\\
\hline

Responsibility and Ethics & Concerns about ethical risks and accountability in ML-enabled systems & (C.8) Lack of Ethical Trust in ML \\
\cline{3-3}
& & (C.14) Lack of Model Accountability \\
\hline
\end{tabular}
\end{table}

\paragraph{\textbf{\underline{Norms, Standards, and Guidelines}}} This group includes two challenges that focus on the lack of standardized metrics and limited integration methods, which can hinder sustainable development.

\begin{description}[leftmargin=0.3cm]

    \item \faSearch\ \textbf{(C.1) Absence of Standardized Metrics \cite{bamiduro2024challenges,habiba2025ml}:} P2, P3, and P4 raised concerns about the absence of standardized sustainability metrics, which make it difficult to measure and compare sustainability efforts.
    
    \smallskip
    \item \faSearch\ \textbf{(C.2) Limited Integration Method:} P1, P2, P5, and P6 pointed out the scarcity of established methods for integrating sustainability into development processes. This lack of guidance may hinder or slow the adoption of sustainable approaches. According to our triangulation process, this challenge has not yet been founded in the scientific literature.
\end{description}

\paragraph{\textbf{\underline{Organizational and Economic Support}}} This group includes three challenges that focus on the lack of awareness, government incentives, and customer demand, which can undermine the motivation and capacity of practitioners to adopt sustainable approaches.
    
\begin{description}[leftmargin=0.3cm]
    \item \faSearch\ \textbf{(C.3) Lack of Awareness \cite{bamiduro2024challenges}}: Participant P2 mentioned a lack of public awareness and publicity surrounding sustainable development efforts. This may contribute to a lack of recognition and motivation for sustainable approaches.

    \smallskip
    \item \faSearch\ \textbf{(C.4) Lack of Government Incentives \cite{bamiduro2024challenges}:} P2, P3, and P5 highlighted the absence of government incentives---both in terms of economic compensation and regulations---to support sustainable projects. P2 said \textit{\say{I see that the companies that look for green things do that for branding ... I think government regulation because has very strict environmental laws, and I think this is driving the companies towards getting more concerned with this, those subjects and then if it is mandatory they cannot avoid it. We need a little bit more laws and compliance for the companies to follow, then they start worrying about this stuff.}}.

    \smallskip
    \item \faSearch\ \textbf{(C.5) Lack of Customer Demand:} Participant P3 and P5 noted the lack of incentives from customers to adopt sustainable developments. A lack of demand for sustainability from clients can reduce the motivation for companies to prioritize sustainability. This challenge represents a novel concern for sustainability research.

\end{description}
\paragraph{\textbf{\underline{Skills and Human Resources}}} This category includes two challenges related to the lack of software frameworks and educational resources that can hinder practitioners’ ability to integrate sustainability considerations in the absence of structured training or supporting infrastructure.
\begin{description}[leftmargin=0.3cm]
    \item \faSearch\ \textbf{(C.6) Lack of Software Frameworks:} All participant highlighted the absence of dedicated frameworks that are specifically engineered to facilitate the incorporation of sustainability considerations into development processes. Also in this case, such a perspective has not yet been discussed by academic literature.

    \smallskip
    \item \faSearch\ \textbf{(C.7) Lack of Educational Resources \cite{tamburri2020sustainable}:} Participants P1, P2, and P4 indicated a lack of educational courses focusing on integrating sustainability into software systems. This education gap may hinder the adoption of sustainable approaches by developers. 
\end{description}

\paragraph{\textbf{\underline{Responsibility and Ethics}}} This category includes one challenge focused on the ethical implications associated with ML-enabled systems for accountability and societal impact.
\begin{description}[leftmargin=0.3cm]
    \item \faSearch\ \textbf{(C.8) Lack of Ethical Trust in ML \cite{pawelec2022deepfakes,kietzmann2020deepfakes}:} Participant P5 described the lack of model trust when it is used and its impact on society; Specifically, P5 reported that: \textit{\say{Then you also have to think about the organization of virtual figures, for example, politicians.}} \textit{\say{Through AI, you generate text that is not there, and many people can fall for it, i.e., they are not all able to be able to understand what is true and what is false. And also the damage can be very heavy, I mean already now with deep fakes and very simple Fake News irreparable damage is created. Imagine fake politicians talking and saying certain things. There is a lot of work to be done on that.}}

\end{description}

\subsubsection{Challenges from the Online Survey}
\label{sec:surveyResults_Approaches_Challenges}
In the survey, respondents were first presented with a list of challenges identified during the interviews and asked to rate their frequency. Table \ref{table:list_challenges} reports these frequencies. As shown, the most significant challenges encountered by ML practitioners were the \textsl{`Lack of government incentives'} and the \textsl{`Lack of established methods'}, followed closely by the \textsl{`Lack of motivation'} and the \textsl{`Lack of software frameworks'}. While other challenges were still considered relevant, they were perceived as less critical. A small proportion of respondents did not identify any significant challenges, with only 6.40\% (13 participants) reporting no difficulties.

Moreover, participants were invited to supplement the list of identified challenges by suggesting additional potential challenges through an open-ended question. These contributions complement the interview findings and map to four different groups in Table~\ref{tab:challenges}, namely: \textsl{`Norms, Standards, and Guidelines'} (\textbf{C.9} and \textbf{C.10}), \textsl{`Organizational and Economic Suppor'} (\textbf{C.11}), \textsl{`Skills and Human Re-
sources'} (\textbf{C.12} and \textbf{C.13}), and \textsl{`Responsibility and Ethics '} (\textbf{C.14}). For five challenges, our triangulation process identified relevant supporting literature, which we reference in the corresponding descriptions to substantiate their contribution to sustainable ML development. However, for one challenge, \textbf{C.13}, we were unable to find the corresponding reference, so we considered it new.


\paragraph{\textbf{\underline{Norms, Standards, and Guidelines}}} This category now includes three additional challenges.
\begin{description}[leftmargin=0.3cm]
    
    \item \faSearch\ \textbf{(C.9) Lack of Sustainability Standard \cite{bamiduro2024challenges,habiba2025ml}:} ML practitioners encounter significant challenges in promoting sustainability within their organizations. The absence of established standards, strategies to reduce emissions, and clear direction adds to the difficulties in promoting sustainable approaches.
    
    \smallskip
    \item \faSearch\ \textbf{(C.10) Sustainability Lacking in Design \cite{bamiduro2024challenges}:} Respondents pointed out that sustainability is an additional requirement to be considered, influencing the overall design of ML-enabled systems and the solutions that are implemented. However, it is often overlooked during the initial stages of model design, making it complicated to integrate sustainable methods at later stages. Practitioners frequently prioritize achieving high performance and meeting specific goals without adequately addressing the ecological implications of their decisions. As a result, systems tend to be resource-intensive and have a significant carbon footprint.
\end{description}

\paragraph{\textbf{\underline{Organizational and Economic Support}}} This category now includes one additional challenge.
\begin{description}[leftmargin=0.3cm]  

    \item \faSearch\ \textbf{(C.11) Lack of Sufficient Financial Support \cite{habiba2025ml}:} Respondents highlighted the challenge of inadequate financial support for sustainability projects. There is a widespread belief that sustainability initiatives can be costly, and the scarcity of essential funds and resources perpetuates this perception. Achieving sustainability goals often requires substantial financial investment, but the necessary resources are often insufficient to support these initiatives. While C.10 and C.11 may seem related, we differentiate them based on their nature. C.10 refers to the lack of time within development workflows to incorporate sustainability approaches, whereas C.11 reflects a lack of funding, institutional prioritization, or incentives. Notably, time constraints may exist even in well-funded projects, while some low-budget projects may still integrate sustainability successfully due to high prioritization.
\end{description}  
\paragraph{\textbf{\underline{Skills and Human Resources}}} This category now includes two additional challenges.
\begin{description}[leftmargin=0.3cm]

    \item \faSearch\ \textbf{(C.12) Lack of Staff Training \cite{tamburri2020sustainable}:} Some participants pointed out a dearth of staff training dedicated to sustainability. Without adequate training and interest, it becomes challenging to instill sustainable values within an organization and equip employees with the skills to drive meaningful change. 

    \smallskip
    \item \faSearch\ \textbf{(C.13) Sustainability Slows Down Software Development:} On a less positive note, some survey responses indicated the presence of certain disadvantages in adopting sustainability approaches during ML development. One significant challenge is the potential slowdown in development time required to ensure effective sustainability improvements. Consequently, there's a need to ensure that software functions correctly within the constraints imposed by sustainable approaches. Furthermore, sustainability-focused models may exhibit lower accuracy and complexity, potentially leading to quality issues in their performance and depth. These drawbacks highlight the need for a careful balance between sustainability considerations and maintaining the desired model quality and effectiveness level. This concern seems not to be discussed in literature, highlighting the need for future studies to investigate how sustainability goals can be balanced with development speed and model quality in practice.
\end{description}    

\paragraph{\textbf{\underline{Responsibility and Ethics}}} This category now includes one additional challenge.
\begin{description}[leftmargin=0.3cm]

    \item \faSearch\ \textbf{(C.14) Lack of Model Accountability \cite{habiba2025ml}:} ML-enabled systems typically require large amounts of data, which often includes sensitive information for training purposes. This dependence can lead to challenges in data storage and management, as well as a lack of metrics and government incentives that address these issues. Additionally, the increasing risk of privacy violations and potential data breaches poses significant concerns. When the data is of poor quality or difficult to quantify, it can undermine decision-making processes, highlighting the importance of responsible data management. It is crucial to establish partnerships that prioritize user comfort, security, and privacy. Furthermore, ML-enabled systems that are not held accountable can potentially harm the population. 
    
\end{description}

\begin{table}
    \centering
    \caption{Frequency of sustainability challenges in ML-enabled systems according to the survey participants.}
    \label{table:list_challenges}
    \footnotesize
    \rowcolors{1}{gray!15}{white}
    {
    \begin{tabular}{|p{5cm}|p{1.5cm}|} \hline 
    \rowcolor{black}
    \textcolor{white}{\textbf{Sustainability Challenge}} & \textcolor{white}{\textbf{Frequency}} \\\hline
    \textit{Lack of Government Incentives} &  \textbf{36.45\% (74)}\\\hline
    \textit{Lack of Methods} &  \textbf{34.98\% (71)}\\\hline
    \textit{Lack of Motivation} &  \textbf{33.00\% (67)}\\\hline
    \textit{Lack of Software Frameworks} &  \textbf{31.53\% (64)}\\\hline
    \textit{Lack of Metrics} &  \textbf{30.54\% (62)}\\\hline
    \textit{Lack of Publicity from Companies} &  \textbf{28.08\% (57)}\\\hline
    \textit{Lack of Customers Intentives} &  \textbf{28.08\% (57)}\\\hline
    \textit{Lack of Courses} &  \textbf{26.11\% (53)}\\\hline
    \textit{Lack of Ethical Trust in ML} &  \textbf{20.20\% (41)}\\\hline
    \textit{Never had problems} &  \textbf{6.40\% (13)}\\\hline
    \end{tabular}}
\end{table}

\definitionbox{RQ$_3$ — Sustainability Challenges}{We identified 14 sustainable challenges, quantified their frequency, and categorized them into four categories. Our analysis reveals that the most common issues include a lack of government incentives, insufficient methodologies, and a lack of motivation. Many respondents highlighted the necessity for increased awareness and training on sustainability approaches, noting that many ML designs prioritize performance over environmental impact or fail to consider sustainability during the initial stages of model design. These findings emphasize the need for a collaborative, multi-stakeholder approach to tackle these challenges, promote best approaches, and ultimately drive the development of more sustainable and ethically responsible ML-enabled systems.}
\section{Discussion and Implications of the Study}
\label{sec:discussion}
Our study provides several implications at different levels and for different stakeholders, from financial organizations to sustainable software engineering. In the following, we discuss the major implications we envision for our research community in an attempt to provide software engineering researchers and practitioners with actionable insights that may contribute to further addressing the sustainability of ML-enabled systems.

\subsection{Positioning our findings within the current state of the art} 
The findings from \textbf{RQ$_1$} shed light on how ML practitioners perceive sustainability, offering a resource for researchers seeking to assess the alignment (or lack thereof) between practitioner views and existing theoretical definitions of sustainability. This comparison reveals both convergence and divergence that can inform future research and guide the development of actionable sustainability frameworks for ML-enabled systems.

In line with prior work \cite{van2021sustainable,mcguire2023sustainability}, practitioners broadly embrace a \emph{socio-technical perspective}, recognizing sustainability as a multifaceted challenge requiring trade-offs between performance, resource efficiency, and ethical responsibility. This aligns with the view of sustainability as a \emph{stratified} and \emph{multisystemic} property spanning technical, organizational, and societal levels \cite{mcguire2023sustainability}.

However, a key misalignment emerges in how sustainability is operationalized. While the literature promotes a lifecycle-wide, strategic integration of sustainability, practitioners often reduce it to a set of short-term, system-level trade-offs, particularly related to energy consumption, computational efficiency, and model size. This suggests that sustainability is frequently treated as an \emph{optimization constraint} rather than a guiding non-functional requirements (NFRs).
This gap points to a broader challenge: to make sustainability actionable, it must be reframed as a measurable NFR, integrated early in the development lifecycle. This need is reinforced by recent industry signals. For instance, Gartner's 2024 press release highlights a growing recognition of software sustainability, predicting that by 2027, 30\% of large global companies will explicitly include sustainability among their NFRs—up from less than 10\% in 2024 \cite{gartner2024}.
Our results contribute to this evolving landscape by capturing how practitioners conceptualize sustainability in real-world settings. These insights can inform the development of pragmatic guidelines, tools, and processes that translate academic principles into approaches tailored to the constraints and priorities of ML development.

\implication{1}{There is a gap between practitioners' focus on operational constraints and academic definitions of sustainability. To bridge this gap, sustainability must be integrated as a practical, measurable NFR throughout the development lifecycle. Industry trends, such as Gartner's recognition of software sustainability, reinforce this need, highlighting a shift toward early adoption of sustainability principles in ML workflows.}

Furthermore, our investigation into \textbf{RQ$_2$} examined the concrete approaches ML practitioners adopt to build sustainable ML-enabled systems. We identified 19 approaches across the three dimensions of sustainability, grouped into five categories. These approaches primarily focus on strategic hardware decisions, individual awareness, and lightweight software strategies. Most approaches relate to environmental sustainability, with an emphasis on energy-efficient hardware and software configurations. Practitioners frequently mention the use of cloud technologies, CPU/GPU optimization, and software techniques to reduce system complexity and size. Perhaps more importantly, these approaches are often applied without dedicated frameworks or tools, highlighting a lack of instrumented support for sustainability-aware ML development.

Our findings align with prior work by Järvenpää et al. \cite{järvenpää2023synthesis}, whose catalog of green architectural tactics also includes hardware optimization (e.g., T24: Use energy-efficient hardware), energy-aware pruning (T15), quantization-aware training (T18), and efficient data management techniques such as removing redundant data (T2) and applying sampling (T1). These approaches contribute directly to lowering computational and energy costs across the ML lifecycle. At the same time, our study reveals additional, underreported approaches, including (A.11) consider adopting cloud technologies from the project's outset, (A.12) consider implementing offline predictions, (A.9) evaluating the suitable model for the task, and (A.6) consider train with synthetic data. These strategies extend the current body of knowledge and can complement the catalog proposed by Järvenpää et al. \cite{järvenpää2023synthesis}.

Regarding social sustainability, our results share similarities with the study by Habiba et al. \cite{habiba2025ml}, particularly in approaches such as (A.15) consider measuring ethical aspects (aligned with P3: Bias and fairness) and (A.19) consider integrating model explainability: (similar to P11: Internal standards). However, our study introduces a distinct focus on (A.4) consider ethical data management, which emphasizes anonymizing sensitive user data prior to model training. Despite these approaches being recognized, our findings highlight a gap between awareness and adoption. Only 24.12\% of survey participants reported actively measuring or addressing fairness in their models. Furthermore, while multiple fairness tools exist, their adoption remains fragmented: 41.38\% of practitioners reported using Microsoft FairLearn, 23.65\% used Google’s What-If Tool, and 20.69\% relied on IBM’s AI Fairness 360, with smaller shares using SHAP (0.98\%) and LIME (0.49\%). These findings point to a lack of systematic guidance and tool integration, underscoring the need for more accessible, unified solutions to support social sustainability in ML workflows.

\implication{2}{ML practitioners of our study primarily focus on hardware efficiency and software optimization. These approaches align with prior research but also introduce new strategies, such as early adoption of cloud technologies and synthetic data usage. However, sustainability efforts are often applied without formal frameworks or tools, highlighting a gap between intention and implementation. This is particularly evident in social sustainability, where fairness evaluations remain limited, and tool adoption is fragmented.} 

\begin{table}[htbp]
    \centering
    \caption{Mapping of Sustainability Approaches to ML-enabled Systems Stages and NFRs. Sustainability Definition (Def).}
    \label{tab:approaches_mapping}
    \footnotesize
    \resizebox{\textwidth}{!}{%
    \rowcolors{2}{gray!15}{white}
    \begin{tabular}{|c|p{4cm}|p{4cm}|p{4cm}|} \hline 
        \rowcolor{black}
        \textcolor{white}{\textbf{Approach}} & 
        \textcolor{white}{\textbf{Description}} & 
        \textcolor{white}{\textbf{Workflow Stage(s)}} & 
        \textcolor{white}{\textbf{NFR(s) Addressed}} \\ \hline

        \textit{Def} & Integrate Sustainability Definition & Model Requirements & Sustainability Dimensions \\ \hline
        \textit{A.13}  & Evaluating the Suitable Model for the Task  & Model Requirements, Model Training  & Cost, Performance, Energy Consumption \\ \hline
        \textit{A.9} & Preferring Simplicity in System Design      & Model Requirements, Model Training, Deployment  & Scalability, Energy Consumption\\ \hline
        \textit{A.4}  & Consider Ethical Data Management            & Data Collection, Data Cleaning  & Privacy, Fairness, Accountability \\ \hline
        \textit{A.5} & Clean and Optimize Dataset & Data Collection, Data Cleaning  & Capacity, Performance, Accuracy, Cost\\ \hline
        \textit{A.6} & Consider Train with Synthetic Data          & Data Collection, Model Training  & Cost, Accuracy, Performance \\ \hline
        \textit{A.11}  & Consider Adopting Cloud Technologies        & Data Collection, Model Training, Deployment  & Energy Efficiency, Cost, Scalability \\ \hline
        \textit{A.1}  & Selecting Efficient Libraries and Languages & Model Training  & Cost, Energy Efficiency, Performance\\ \hline
        \textit{A.2}  & Optimizing Code to Minimize Resources         & Model Training  &  Cost, Energy Efficiency, Performance\\ \hline
        \textit{A.7}  & Consider Adopting Pre-trained Models          & Model Training  & Cost, Energy Efficiency, Accuracy, Transferability\\ \hline
        \textit{A.8}  & Consider Streamlining Models for Efficiency & Model Training  & Cost, Energy Efficiency, Capacity, Scalability \\ \hline
        \textit{A.10} & Consider Delegating Learning to Exploration & Model Training  & Cost, Performance\\ \hline
        \textit{A.17} & Consider Knowledge Distillation             & Model Training  & Capacity, Cost, Energy Efficiency, Performance \\ \hline
        \textit{A.11} & Adopt Strategies for Efficient Management of ML Costs & Model Training, Monitoring  & Cost, Energy Efficiency \\ \hline
        \textit{A.15}  & Consider Measuring Ethical Aspects           & Model Evaluation  & Fairness, Accountability\\ \hline
        \textit{A.19} & Consider Integrate Model Explainability & Model Evaluation, Monitoring  & Interpretability, Transparency, Explainability \\ \hline
        \textit{A.12}  & Consider Implementing Offline Predictions & Deployment  & Cost, Energy Efficiency, Performance \\ \hline
        \textit{A.14} & Operational Sustainability in AI Devices       & Deployment  & Capacity, Energy Efficiency \\ \hline
        \textit{A.18} & Consider New Hardware                          & Deployment  & Cost, Energy Efficiency, Performance \\ \hline
        \textit{A.16} & Track Sustainability Metrics & Model Monitoring  & Energy Efficiency, Cost, Fairness, Accuracy\\ \hline
    \end{tabular}%
    }
\end{table}

Finally, our study adds a novel contribution to the literature by surfacing approaches and challenges related to economic sustainability in ML-enabled systems - this is a dimension that, to our knowledge, has not been systematically addressed in prior empirical work. Our findings show that sustainability in ML is not purely a technical matter to be solved through algorithmic or hardware optimizations. Instead, it is a socio-technical concern, shaped by industry constraints, evolving regulations, and the availability of tools that support sustainable approaches. Economic sustainability plays a critical role in real-world ML adoption, where development costs, resource allocation, and budget limitations directly influence decision-making. Practitioners must balance performance with cost-efficiency, navigating trade-offs among infrastructure investments, computational needs, and long-term maintainability. Regulatory pressures, like energy usage standards and ethical AI guidelines, further compound these challenges, introducing new costs but also spurring innovation in sustainable approaches. By foregrounding economic considerations, our study complements prior work that has focused predominantly on environmental and social aspects. This expanded lens reinforces the need for integrated strategies that address technical feasibility, regulatory compliance, and financial viability across the ML development lifecycle.

\implication{3}{Our findings reveal that sustainability in ML-enabled systems is a socio-technical issue influenced by industry constraints, regulations, and financial feasibility. While prior research has focused on environmental and social sustainability, we emphasize the need to incorporate economic sustainability, particularly in reducing development costs and ensuring long-term maintainability.} 

\subsection{Using the Sustainable Approaches Elicited to Optimize Non-Functional Requirements}
To make our contribution more practical, we performed an additional step by mapping the sustainability approaches identified in \textbf{RQ$_2$} to both the phases of the ML development workflow and the specific NFRs they address.
Following the nine-stage ML workflow proposed by Amershi et al. \cite{amershi2019software}, we systematically positioned each sustainability practice based on when participants reported applying it, from model requirements to deployment and monitoring. This mapping was conducted by the first author and drew on methodologies used in prior studies \cite{järvenpää2023synthesis,habiba2025ml}. We then linked each approach to relevant NFRs using the framework by De Martino and Palomba \cite{MARTINO2025107678}, analyzing both the motivations and intended impacts described by participants. For example, the approach ``(A.6) Consider Training with Synthetic Data'' was commonly cited as a way to reduce data acquisition costs and improve training efficiency. Based on participant feedback, we mapped this approach to the ``Data Collection'' and ``Model Training'' phases and associated it with the NFRs of cost, accuracy, and performance.
The results of this analysis are shown in Table \ref{tab:approaches_mapping}, which shows how sustainability considerations are distributed across all ML workflow stages. The mapping also highlights the diverse range of NFRs addressed. For instance, energy and cost-related approaches include ``(A.11) Consider Adopt Cloud Technologies'' and ``(A.12) Consider Implement Offline Predictions'', while socially oriented approaches, such as ``(A.4) Consider Ethical Data Management'' and ``(A.15) Consider Measure Ethical Aspects'', focus on fairness and equity. Other approaches, such as consider knowledge distillation (A.17) and consider integrate model explainability (A.19), contribute to explainability and scalability, underscoring the multifaceted nature of sustainability in ML-enabled systems.
Interestingly, several approaches concentrate around the model training phase, including resource-efficient training, pre-trained model reuse, and library selection. In contrast, relatively few approaches target post-deployment phases like monitoring—despite its importance for tracking sustainability over time. Approaches such as ``(A.16) track sustainability metrics'' hint at this gap, suggesting opportunities for future work to develop tools and methods that support sustainability beyond initial deployment. Overall, this structured mapping offers a practical framework for understanding how sustainability approaches intersect with both ML workflows and NFRs. It supports researchers and practitioners in systematically integrating sustainability from the outset—moving beyond isolated or reactive efforts toward a more holistic, lifecycle-aware approach.

\implication{4}{The mapping confirms that sustainability should be integrated across the entire ML lifecycle. While many approaches focus on optimizing resources during model training, post-deployment monitoring remains underexplored. The variety of addressed NFRs further highlights the multifaceted nature of sustainability in ML-enabled systems.}

\subsection{Breaking the barriers to the adoption of sustainability approaches}
Despite increasing awareness, our findings reveal several persistent barriers hindering the effective adoption of sustainability in ML-enabled systems. These include the absence of standardized metrics, limited organizational support, and inadequate tool support, contributing to inconsistent approaches across ML workflows.
A major challenge is the lack of standardized sustainability metrics and evaluation frameworks. Without clear benchmarks, practitioners struggle to measure and compare the impact of sustainability approaches, leading to fragmented adoption. This reflects broader challenges in software engineering \cite{bamiduro2024challenges}, and highlights the need for domain-specific, actionable indicators to guide practice and foster accountability. Limited motivation and awareness also emerged as key barriers. Performance, cost, and time-to-market often take precedence over sustainability, especially in the absence of organizational incentives or regulatory pressure. Consequently, addressing these issues requires both cultural change and structural incentives, such as policies, certifications, or financial benefits for sustainable development. This represents a \emph{call for research actions} aimed at identifying effective mechanisms to embed sustainability into standard ML development approaches. Further studies are needed to explore how sustainability can be quantified, benchmarked, and integrated without compromising system performance or development efficiency. This includes developing practical assessment frameworks, investigating the trade-offs between sustainability and other quality attributes, and proposing automated tools that assist practitioners in making sustainability-aware decisions throughout the ML lifecycle. 

In addition to conceptual challenges, practical constraints like inadequate funding and limited tooling hinder adoption. Many sustainability approaches require added computational resources or specialized hardware, increasing development costs. Practitioners also report a lack of integrated tools for sustainability evaluation and monitoring. Research can help by designing cost-effective, lightweight, and automated tools that integrate into existing ML workflows—ideally as open-source frameworks to lower entry barriers. Promising directions include AI-driven optimization to balance sustainability and performance, and economic studies quantifying long-term cost and energy savings. Finally, training and education gaps remain. Few practitioners have formal instruction in sustainable ML, leading to ad hoc or superficial efforts. Expanding curricula and offering industry training on sustainability principles could help foster best practices from early career stages.

\implication{5}{Sustainability adoption in ML-enabled systems is limited by the absence of standardized metrics, integration methods, organizational support, and practical tools. Overcoming these barriers requires clear metrics, automated and cost-effective tools, supportive policies, and incentives. Assessing economic trade-offs and expanding education and training are also key to promoting long-term, sustainable approaches in ML.}
\section{Threats to validity}
\label{sec:threats}
A number of design decisions might have induced threats to the validity of the study. In the following, we discuss and elaborate on how we mitigated them. 

\subsection{Construct Validity} These limitations regard imprecision in performed measurement. In order to enhance the validity of measurement and ascertain the clarity, impartiality, and alignment with research objectives, we employed Dillman's and Ciolkowski et al. guidelines \cite{dillman2014internet, ciolkowski2003practical}. These guidelines, widely recognized and applied by experts in the field, helped us create the process for preparing, conducting, and analyzing all phases of the study.

The study's main objective was to collect insight into sustainability in ML-enabled systems. For such a reason, the way we defined sustainability—if not correct—could lead to unreliable results. To avoid this, we introduce Brundtland's sustainability definition \cite{brundtland1987our} and then describe the dimensions of sustainability in the software process of McGuire et al. \cite{mcguire2023sustainability}. Another potential threat to construct validity relates to how we asked participants to rate the importance of sustainability dimensions. Direct rating questions may introduce social desirability bias or acquiescence \cite{grimm2010social}, as participants might feel compelled to rate all sustainability dimensions as important to align with perceived expectations or to avoid appearing dismissive of socially valued concerns. To mitigate this concern, we first asked participants to define sustainability in their own words, followed by both Likert-scale assessments and a ranking question. The ranking question helped reduce bias by prompting explicit prioritization, thereby improving the robustness of our measurements.

As a final note, to ensure the study's robustness, we conducted two pilot studies involving eight ML practitioners: four participated in piloting the interviews, and four others piloted the survey. These pilot tests were essential in identifying potential errors and flaws, which were promptly addressed and corrected, ensuring the validity and reliability of the two phases of the study.

\subsection{External Validity} These limitations concern the generalizability and transferability of the findings obtained. In the first phase of the qualitative study, we contacted 35 ML practitioners, but only eight ML practitioners agreed to be interviewed, a small number. Participants worked in specific working environments (e.g., economic and political domains); our results could be highly related to these contexts. In order to mitigate all of the above limitations and include a larger group to mitigate the generalizability of the results, we conducted a second phase of the quantitative study by including a larger population (203 participants); specifically, we conducted a large-scale survey study, to make the results more robust; this practice is also used in similar published work in the software engineering domain~\cite{pant2023_ethics_GT, sayagh2018software, manotas2016empirical, leitner2019mixed}. For our survey study, we performed a pre-screening step to identify many individuals with heterogeneous characteristics (e.g., academic background and knowledge). In the end, our findings are quantified in the survey. 

\subsection{Internal Validity} 
These limitations pertain to unforeseen factors impacting the results of our study. Qualitative data analysis is an activity that depends primarily on the skills and perceptions of researchers. For this reason, systematic approaches—like the one of grounded theory—arose to support researchers and led to more robust and recognizable findings. To avoid misunderstandings and misinterpretations of qualitative data, we applied the approach chosen for the analysis. Moreover, three authors independently performed the analysis, later computing the Cohen’s Kappa~\cite{cohen1960coefficient} inter-rater agreement. Furthermore, the final list of categories was refined through a series of meetings that involved all the authors of the paper, who all contributed to making the conclusions drawn more reliable.

\subsection{Conclusion Validity} These limitations concern the reliability of the conclusions drawn from the data. In the qualitative study, we collected initial insight from eight ML practitioners. However, to mitigate the risk of drawing conclusions based on limited perspectives, we conducted a quantitative survey with 203 participants to validate and extend our findings. Additionally, to ensure the inclusion of possible additional findings and to capture potentially overlooked insights, we incorporated open-ended survey questions. This allowed participants to contribute additional results beyond the default response, ensuring that our results covered a greater number of items that practitioners felt were relevant. Lastly, a potential threat to conclusion validity in our study is the presence of AI-generated responses among participants, which could introduce biases or distortions in our findings. To mitigate this risk, we employed the ZeroGPT tool to remove AI-generated texts.
\section{Conclusion}
\label{sec:conclusion}
In this paper, we aimed at understanding the ML engineer's perspective on sustainability within ML-enabled systems. Our work provided the following contributions:

\begin{enumerate}
    \item How sustainability is perceived in practice, what are the practices and strategies employed by ML practitioners to deal with sustainability concerns, and what are the current challenges that ML practitioners face;

    \smallskip
    \item A set of data, findings, and implications that may drive the future software engineering efforts on the matter toward improved methods and tools to design sustainable ML-enabled systems;

    \smallskip
    \item A public replication package \cite{appendix} which may be exploited by researchers to either replicate our work or build on top of our findings, e.g., by means of additional socio-technical grounded theory analysis \cite{9520216} on the answers provided by the practitioners involved.
    
\end{enumerate}

Our next research will focus on developing nudge strategies to assist in software engineering, with a specific emphasis on leveraging behavioral theory \cite{lehner2016nudging,klaniecki2016behaviour}. This approach involves creating subtle but impactful methodologies that guide software engineers toward more sustainable decision-making practices and processes. By understanding and applying the principles of behavioral science, we aim to design interventions that influence practitioners' behaviors and choices in a positive direction, promoting sustainability both directly and indirectly in software engineering. At the same time, we aim to design novel software engineering tools and techniques to address sustainability challenges.
Moreover, it is important to note that the practices and opinions reported by our participants should not be considered as reference standards. Practitioners are not infallible, and local constraints, misunderstandings, or personal biases may influence their approaches. Thus, a key implication for future work is the need for further empirical research to evaluate these practices in terms of their effectiveness, impact, and broader applicability.

\section*{Acknowledgments}
ChatGPT was used to improve the language and readability of this work. The authors reviewed and edited the content and take full responsibility for it. This work has been partially supported by the Grant PID2024-156019OB-I00 funded by MICIU/AEI/10.13039/501100011033 by ERDF, EU, and \textsl{GAISSA-Optimizer} research project funded by the AGAUR agency (Code: 2025 PROD 00236). This work has been partially supported by the \textsl{Qual-AI} and \textsl{FRINGE} national research projects funded by the MUR under the PRIN 2022 and PRIN 2022 PNRR programs (Codes: D53D23008570006 and D53D23017340001, respectively). 

\bibliographystyle{ACM-Reference-Format}
\bibliography{bib/bib}

\end{document}